\documentclass{emulateapj}
\usepackage{apjfonts,graphicx,lscape,subfigure,natbib}
\slugcomment{Received 2011 April 4; accepted 2011 October 25; to appear in ApJ}

\newcommand{\angstrom}{{\rm \AA}}

\newcommand{\hseventy}{h{$_{70}^{-1}$}}
\newcommand{\hbeta}{H{$\beta$}}
\newcommand{\halpha}{H{$\alpha$}}
\newcommand{\hdelta}{H{$\delta$}}
\newcommand{\hdeltaa}{H{$\delta_{{\rm A}}$}}
\newcommand{\dnfk}{$D_n(4000)$}

\newcommand{\OII}{[O{\sevenrm\,II}]}
\newcommand{\OIII}{[O{\sevenrm\,III}]}
\newcommand{\loiii}{$L_{{\rm [O\,III]}}$}

\newcommand{\OIIIb}{[O{\sevenrm\,III}]\,$\lambda$5007}

\newcommand{\NII}{[N{\sevenrm\,II}]}
\newcommand{\NIIb}{[N{\sevenrm\,II}]\,$\lambda$6584}

 \font\sevenrm=cmr7 scaled 1000

\begin{document}

\title{Active Galactic Nucleus Pairs from the Sloan Digital Sky Survey. \\
II. Evidence for Tidally Enhanced Star Formation and Black Hole
Accretion}

\shorttitle{AGN Pairs in SDSS. II}

\shortauthors{Liu, Shen, \& Strauss}
\author{Xin Liu\altaffilmark{1,2,3}, Yue Shen\altaffilmark{1},
and Michael A. Strauss\altaffilmark{2}}

\altaffiltext{1}{Harvard-Smithsonian Center for Astrophysics,
60 Garden St., Cambridge, MA 02138}

\altaffiltext{2}{Department of Astrophysical Sciences,
Princeton University, Peyton Hall -- Ivy Lane, Princeton, NJ
08544}

\altaffiltext{3}{Einstein Fellow}

\begin{abstract}
Active galactic nuclei (AGNs) are occasionally seen in pairs,
suggesting that tidal encounters are responsible for the
accretion of material by both central supermassive black holes
(BHs). In Paper I of this series, we selected a sample of AGN
pairs with projected separations $r_p <100$ \hseventy\ kpc and
velocity offsets $<600$ km s$^{-1}$ from the SDSS DR7 and
quantified their frequency. In this paper, we address the
BH-accretion and recent star-formation properties in their host
galaxies.  AGN pairs experience stronger BH accretion, as
measured by their \OIIIb\ luminosities (corrected for
contribution from star formation) and Eddington ratios, than do
control samples of single AGNs matched in redshift and
host-galaxy stellar mass. Their host galaxies have stronger
post-starburst activity and younger mean stellar ages, as
indicated by stronger \hdelta\ absorption and smaller 4000
\angstrom\ break in their spectra. The BH accretion and recent
star formation in the host galaxies both increase with
decreasing projected separation in AGN pairs, for $r_p \lesssim
10$--$30$ \hseventy\ kpc. The intensity of BH accretion, the
post-starburst strength, and the mean stellar ages are
correlated between the two AGNs in a pair.  The luminosities
and Eddington ratios of AGN pairs are correlated with recent
star formation in their host galaxies, with a scaling relation
consistent with that observed in single AGNs.  Our results
suggest that galaxy tidal interactions enhance both BH
accretion and host-galaxy star formation in AGN pairs, even
though the majority of low redshift AGNs is not coincident with
on-going interactions.
\end{abstract}


\section{Introduction}\label{sec:intro}

It has long been recognized that galaxy mergers and close
interactions play a vital role in shaping the morphology of
galaxies
\citep[e.g.,][]{zwicky56,vorontsov59,toomre72,khachikian87,sanders88,barnes92}
and in particular the cores of ellipticals
\citep[e.g.,][]{faber97,graham04,merritt04,lauer05,kormendy09}.
While mergers only represent brief episodes in the lifetime of
a galaxy \citep[e.g.,][]{lotz08,conselice09,darg10a,lotz11},
there is circumstantial evidence that they can trigger
substantial star formation both in the nuclear region and
throughout the galaxy \citep[e.g., see the review
by][]{struck06}. The enhancement in star formation predicted by
numerical simulations depends on how that star formation and
its associated feedback are modeled
\citep[e.g.,][]{mihos92,kauffmann93,cole94,springel03,springel05}.
Therefore observations of star formation in interacting
galaxies can be used to constrain these models, and help
advance our general understanding of star formation, and in
particular, modes that are exclusive to strong tidal encounters
\citep[e.g.,][]{elmegreen93,daddi10} which may be more common
in early galaxy formation \citep[e.g.,][]{forster09}.

Galaxy mergers and close interactions are also thought to drive
significant accretion in their central supermassive black holes
\citep[BHs; e.g.,][]{hernquist89,moore96} and are thus
responsible for at least some Active Galactic Nuclei (AGNs).
The enhanced nuclear activity may be primarily driven by tidal
torques, or secondarily fueled by stellar mass loss due to
elevated levels of star formation
\citep[e.g.,][]{norman88,ciotti07,ciotti10}. Some numerical
models predict that merger-induced AGNs heat the gas
surrounding them, preventing further gas cooling and star
formation
\citep[e.g.,][]{springel05,springel05b,dimatteo05,hopkins06}.
Observationally, however, the causal link between galaxy
interactions and AGNs has been controversial. The details of
AGN feeding, and in particular, the role of mergers, may be
quite different at high and low accretion rates
\citep[e.g.,][]{ho08}. While the small-scale ($\sim0.1$--1 Mpc)
quasar-quasar two-point correlation function suggests a
clustering excess over the large-scale ($>1$ Mpc) extrapolation
\citep{hennawi06,myer07,hennawi09,shen10}, it is unclear
whether this is due to interactions of the quasar pairs' host
galaxies \citep[e.g.,][]{djorgovski91,kochanek99,mortlock99},
or is rather a manifest of the small-scale clustering strength
of their host dark matter halos \citep[e.g.,][]{hopkins08}.
Some observations found an excess of close neighbors in AGN
host galaxies or a higher fraction of AGNs in interacting than
in isolated galaxies
\citep[e.g.,][]{petrosian82,hutchings83,dahari84,kennicutt84,keel85,bahcall97,serber06,koss10,silverman11,ellison11}
whereas others detected no significant difference
\citep[e.g.,][]{dahari85,schmitt01,miller03,grogin05,waskett05,ellison08,li08b,darg10a}.

\begin{figure*}
  \centering
    \includegraphics[width=85mm]{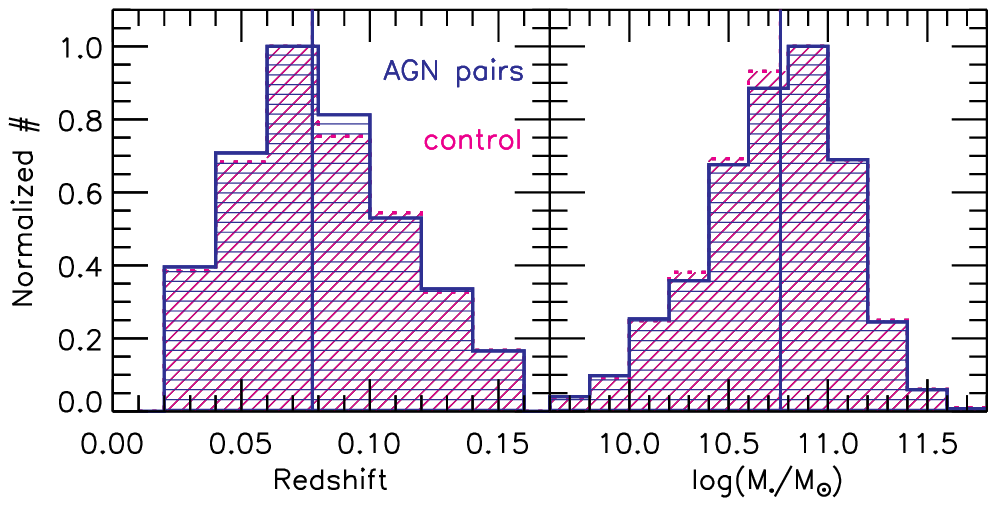}\qquad
    \includegraphics[width=85mm]{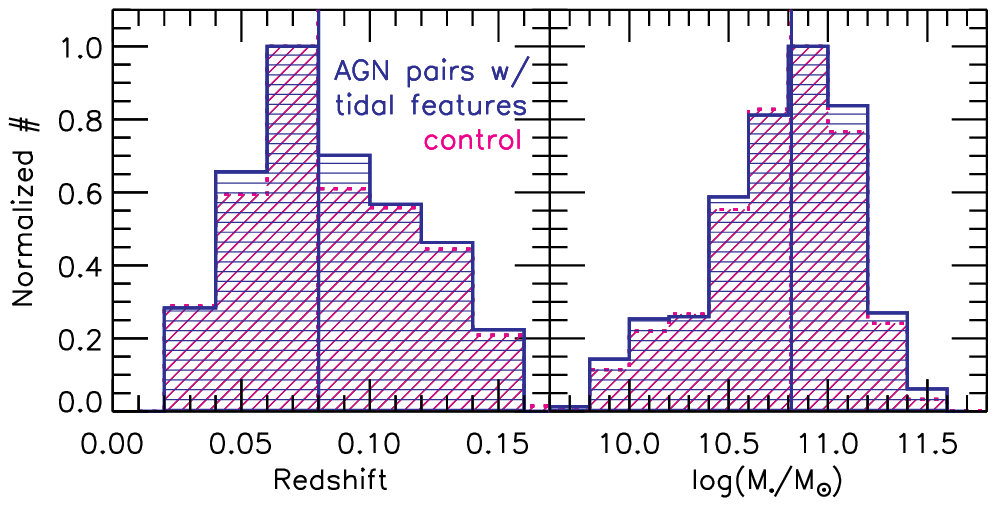}
    \caption{Redshift and stellar mass distributions
    of AGN pairs (left panel; blue) and the subset with
    tidal features (right panel; blue) compared
    to their control samples (magenta in both panels).
    The AGN pair and control samples are matched
    in redshift and stellar mass (with KS probabilities larger than 50\%).
    The median value of each distribution is indicated with a vertical line.}
    \label{fig:control}
\end{figure*}

Observations of galaxy pairs directly probe the effects induced
by galaxy tidal interactions, at least at the early stages when
the two galaxies can be identified as separate. Studies of the
properties of galaxy pairs as a function of pair separation,
and in comparison to single galaxies provide clues to how tidal
effects evolve as mergers progress. Simulations suggest that
merger-induced star formation and AGN activity peak after the
galaxies have coalesced \citep[][]{springel05,dimatteo05}.
There is accumulating evidence for tidally-enhanced activity
starting at the early merger stages when the galaxies are still
separate and that the enhancement becomes more prominent as a
merger progresses. Emission-line galaxies in pairs show
stronger star formation activity, which is larger for smaller
projected separations
\citep{barton00,lambas03,alonso04,nikolic04,ellison08}.
Galaxy-AGN pairs with smaller separations show stronger AGN
\citep{rogers09}, while galaxy pairs with smaller separations
host a higher fraction of AGN \citep{woods07,darg10b}. If tidal
effects impact both interacting components (e.g., by promoting
gas infall and inducing gravitational instabilities) and the
induced activity is synchronous, there should be observable
correlations between them, unless additional factors regulate
such trends. Indeed, \citet{holmberg58} remarked on what is now
called the ``Holmberg effect'', whereby the colors of galaxies
in pairs are correlated \citep[e.g.,
][]{madore86,kennicutt87,laurikainen89} which suggests the
presence of tidally induced star formation in both of the
interacting galaxies.

We have selected a sample of 1286 pairs of AGNs with projected
separations $r_p <100$ \hseventy\ kpc and line-of-sight
velocity offsets $\Delta v <600$ km s$^{-1}$ from the Seventh
Data Release \citep[DR7;][]{SDSSDR7} of the Sloan Digital Sky
Survey \citep[SDSS;][]{york00} and examined their frequency
among the parent sample of optically selected AGNs
\citep[][Paper I]{liu11a}. The sample is defined in detail in
Paper I. In the present paper, we study their BH-accretion and
recent star-formation properties in the host galaxies to
constrain the effects of galaxy tidal interactions.  AGN pairs
characterize a special population of galaxy pairs, in which the
central SMBHs of both galaxies are actively accreting at the
same time. A statistical sample of AGN pairs enables us to
examine whether there is a correlation in the accretion power
of the central SMBHs between the interacting components, akin
to what has been observed for star formation in galaxy pairs.

In \S \ref{sec:data} we describe the construction of our AGN
pair sample and the control sample of single AGNs.  In \S
\ref{sec:result} we present the recent star formation and BH
accretion properties of AGN pairs, and we examine their
dependence on the projected pair separation and recessional
velocity offset in \S \ref{subsec:prop}, the dependence on
host-galaxy properties in \S \ref{subsec:progenitor}, the
correlation between the two components in each AGN pair in \S
\ref{subsec:correlation}, and the correlation between recent
star formation and AGN activity in \S \ref{subsec:sfagn}.  We
discuss the implications of our results in \S \ref{sec:discuss}
and conclude in \S \ref{sec:sum}. Throughout we assume a
cosmology with $\Omega_m = 0.3$, $\Omega_{\Lambda} = 0.7$, and
$H_{0} = 70$ $h_{70}$ km s$^{-1}$ Mpc$^{-1}$.

\begin{figure*}
  \centering
    \includegraphics[width=85mm]{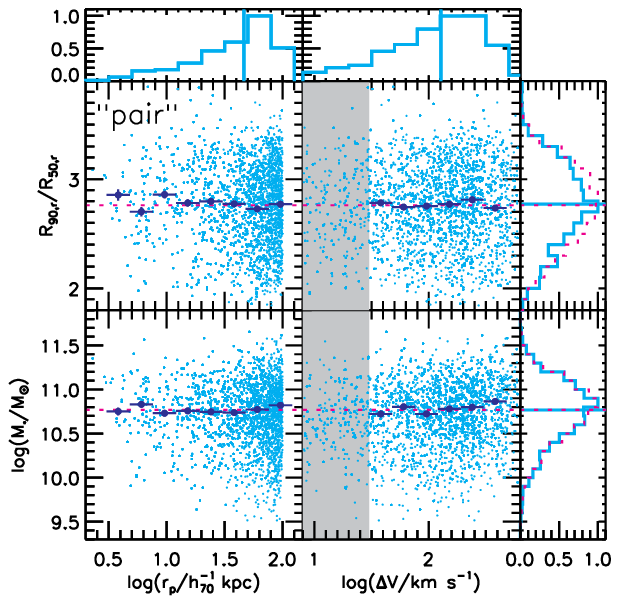}\qquad
    \includegraphics[width=85mm]{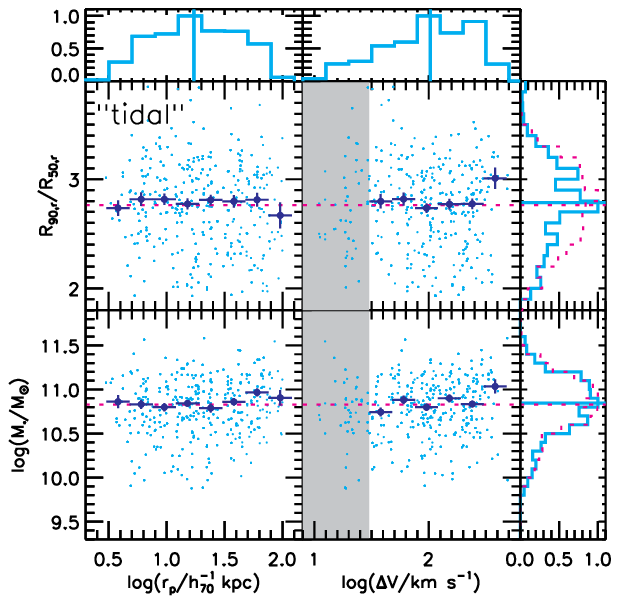}
    \caption{Stellar mass $M_{\ast}$ and $r$-band concentration index
    $R_{90,r}/R_{50,r}$ as functions of
    interaction parameters $r_p$ and $\Delta v$ for the pair (left block) and
    the tidal (right block) samples.
    Here and throughout, we plot both galaxies in each AGN pair.
    Cyan dots denote individual objects, whereas blue filled circles
    represent median values at each bin of $r_p$ or $\Delta v$.
    Error bars indicate uncertainties of the median values.
    The histograms plotted in thick curves in cyan are normalized distributions
    of AGN pairs corrected for SDSS spectroscopic incompleteness
    due to fiber collisions (Paper I).
    The thin dotted curves in magenta are for the control AGNs
    matched in redshift and in stellar mass (Figure \ref{fig:control}).
    Vertical and horizontal lines denote median values of the distributions.
    Here and throughout, we do not plot the median values below $\Delta v = 30$ km/s
    [i.e., log$(\Delta v$/km s$^{-1})< 1.5$] which are insignificant because
    they are dominated by redshift measurement uncertainties.}
    \label{fig:control_at_fixed_r}
\end{figure*}

\section{Data: AGN Pairs, AGN Pairs with Tidal Features, and Control
Samples}\label{sec:data}

We briefly describe the basic properties and physical
measurements of our AGN pairs. Details of our sample selection
are presented in Paper I.  We have selected AGN pairs with $r_p
<100$ \hseventy\ kpc and $\Delta v <600$ km s$^{-1}$ from a
parent sample of 138,070 AGNs spectroscopically identified
based on their optical diagnostic emission-line
ratios\footnote{For narrow emission line objects, we adopt the
\citet{kauffmann03} empirical criterion for AGN classification
as our baseline values (i.e., including AGN-H {\tiny II}
composites to avoid selection biases against galaxies with more
starburst components), but we have also tested that our results
do not vary qualitatively using the \citet{kewley01}
theoretical ``starburst limit'' criterion (i.e., excluding
AGN-H {\tiny II} composites to focus on AGN dominated
systems).} and/or widths. 98\% of the sample is contained in
the SDSS main galaxy catalog \citep{strauss02}, and for the
present analysis we have excluded broad-line objects
\citep{hao05a} to avoid AGN continuum contamination on
host-galaxy measurements. We refer to this sample as the
``pair'' sample. As discussed in Paper I, 256 of the 1286 AGN
pairs show unambiguous morphological tidal features tails in
the SDSS images, such as bridges, and/or rings, indicative of
strong tidal interactions, and we refer to this subset as the
``tidal'' sample.  We consider the tidal sample to be
``cleaner'' than the pair sample for studies of interactions,
as the pair sample may include closely separated AGN pairs that
are not yet tidally interacting. On the other hand, as
discussed in Paper I, the tidal sample is incomplete and
subject to biases due to the surface-brightness and
image-resolution limits in our ability to recognize tidal
features.  We thus analyze both samples and compare results to
quantify the range of possible values. As discussed in Paper I,
due to the finite size of the SDSS fibers, galaxy pairs with
separations smaller than 55\arcsec\ will not both be observed
unless they fall in the overlap regions on adjacent plates
\citep{strauss02,blanton03}.  We correct for this spectroscopic
incompleteness by supplementing the observed sample of AGN
pairs with $(C-1)N_{p}$ repeated systems randomly drawn from
the AGN pairs with separations smaller than 55\arcsec , where
$N_{p}$ is the observed number of pairs and $C\approx 3.3$ is
the correction factor (see Paper I for details). The small
separation AGN pairs fall in the overlap regions on adjacent
plates so that they both got spectroscopic observations. There
should be no biases introduced here because the
spectroscopically observed small separation pairs are randomly
distributed on the sky and are therefore representative of the
parent small separation population.

We adopt redshifts $z$ and stellar velocity dispersions
$\sigma_{\ast}$ of each galaxy from the \texttt{specBS}
pipeline (\citealt{SDSSDR6}; \citealt{SDSSDR8}; see also
discussion in \citealt{blanton05}) and excluded extreme values
of $\sigma_{\ast}$ ($\sigma_{\ast} <30$ or $>500$ km s$^{-1}$)
and those with large uncertainties (S/N$<3$) or negative errors
from bad fits. Additional spectral and photometric properties
such as emission-line fluxes, continuum spectral indices,
stellar masses, and half-light radii are taken from the MPA-JHU
data product\footnote{http://www.mpa-garching.mpg.de/SDSS/DR7/}
(see \citealt{SDSSDR8} for a description of the catalog). The
emission-line measurements are from Gaussian fits to
continuum-subtracted spectra \citep{brinchmann04,tremonti04}.
The emission-line fluxes have been normalized to SDSS $r$-band
photometric fiber magnitudes and have been corrected for
Galactic foreground extinction following \citet{odonnell94}
using the map of \citet{schlegel98}.  Continuum spectral
indices such as \hdeltaa\ and \dnfk\ are measured from data
spectra after subtracting all 3-$\sigma$ emission lines
\citep{kauffmann03c,brinchmann04}, which, as we have checked,
are consistent with those measured from best-fit continuum
model spectra. The stellar mass estimates are total stellar
masses derived from population synthesis fits using the
\citet{bc03} models to SDSS broad-band photometry
\citep{kauffmann03c,salim07}. The adopted stellar mass estimate
from photometry fits is a good indicator of the dynamical mass
inside the effective radius ($M_{{\rm dyn}} \propto
\sigma_{\ast}^2 R_e/G$, where $R_e$ is the galaxy effective
radius and $G$ is the gravitational constant) for $M_{\ast} >
10^{10} M_{\odot}$ \citep{brinchmann00,drory04,padmanabhan04};
At the lower mass end of our sample ($M_{\ast} \sim 10^{9.5}
M_{\odot}$), $M_{\ast}$ is $\sim 0.2$ dex larger than $M_{{\rm
dyn}}$ \citep{drory04}.

To characterize how the effects of tidal interactions on host
star formation and BH accretion evolve in AGN pairs as the
merger progresses, we examine their statistical properties as a
function of pair separation and velocity offset, and compare to
control samples of single AGNs.  We draw control AGNs from our
parent sample of 138,070 AGNs. We match the control sample to
the AGN pair sample in redshift to mitigate selection biases in
our flux-limited sample, and aperture bias due to the
difference in the physical scales covered by the SDSS 3\arcsec\
fibers for galaxies at different redshifts.  We also match
stellar mass distribution to control mass-dependent effects of
AGN host-galaxy properties \citep{kauffmann03}. We show in
Figure \ref{fig:control} the redshift and stellar mass
distributions of the pair (tidal) sample and its control
sample, respectively. We match the whole distribution rather
than object by object. The control samples are drawn to have
the same distribution in redshift and in stellar mass as the
pair (tidal) sample by requiring that their Kolmogorov-Smirnov
(KS) probabilities are larger than 50\% at least. We have
matched galaxy mass rather than luminosity in our control
sample, because interacting galaxies often show younger stellar
populations than do isolated galaxies, making galaxy luminosity
a biased indicator of mass \citep[see also e.g.,][]{ellison08}.

\begin{figure*}
  \centering
    \includegraphics[width=85mm]{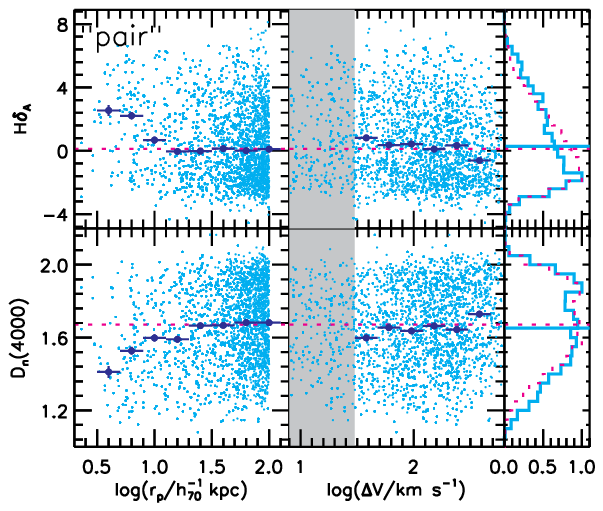}\qquad
    \includegraphics[width=85mm]{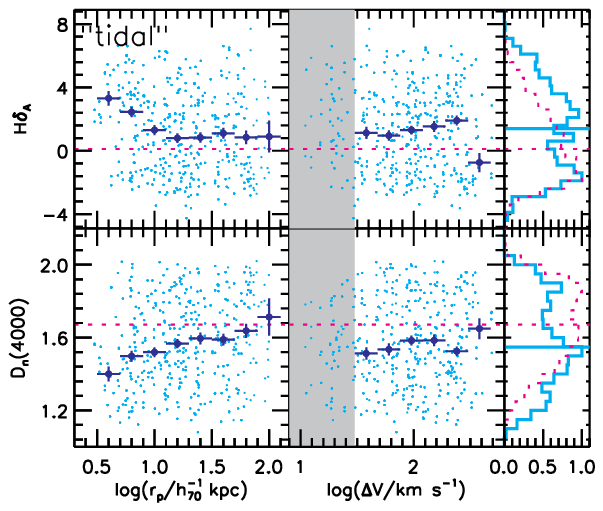}
    \caption{Spectral indices \hdeltaa\ and \dnfk\ as
    functions of interaction parameters for the pair (left block) and
    tidal (right block) samples.
    Symbols and line types are the same as
    in Figure \ref{fig:control_at_fixed_r}.
    }
    \label{fig:sf}
\end{figure*}

To study the intrinsic dependence of host star formation and BH
accretion on $r_p$ and $\Delta v$ in AGN pairs, we first verify
in Figure \ref{fig:control_at_fixed_r} that host-galaxy stellar
mass does not correlate with $r_p$ or $\Delta v$, obviating the
need to match control samples for each subset of AGN pairs at
any fixed $r_p$ or $\Delta v$.  Figure
\ref{fig:control_at_fixed_r} also shows that galaxy
concentration, a measure of the mass distribution within
galaxies, does not correlate with $r_p$ or $\Delta v$, which is
relevant because the distribution of mass within galaxies may
also regulate merger timescales\footnote{The mass distribution
of dark matter halos at large radius dictates merger timescales
on tens-of-kpc scales.}. The $r$-band concentration index $C_r$
is defined as the ratio between the $r$-band Petrosian
half-light radius and the radius enclosing 90\% of the $r$-band
Petrosian luminosity of a galaxy \citep{strauss02}.

\section{Results: Black Hole Accretion and Host-Galaxy Recent Star Formation in AGN
Pairs}\label{sec:result}

In \S \ref{subsec:prop} we present the distributions of BH
accretion and host recent star formation properties of the pair
and tidal samples and their dependence on $r_p$ and $\Delta v$,
and compare them with control AGN samples. We show how these
results depend on host-galaxy stellar mass ratio and
concentration in \S \ref{subsec:progenitor}. \S
\ref{subsec:correlation} focuses on the correlations between
the interacting components in an AGN pair. We then examine in
\S \ref{subsec:sfagn} the correlation between BH accretion and
host recent star formation in AGN pairs and compare with those
of control samples of single AGNs.

\subsection{Dependence on Projected Pair Separation and Recessional Velocity
Offset}\label{subsec:prop}

\subsubsection{Recent Star Formation in the Host Galaxies}\label{subsubsec:sf}

In this section, we address how galaxy tidal encounters affect
recent star formation in AGN pairs.  We cannot directly use the
commonly used optical emission lines (\halpha\ and \OII ) as
indicators of current star formation for the galaxies in our
sample because they contain contribution from AGN
\citep[e.g.,][]{ho97,kauffmann03,brinchmann04}. Instead we
adopt the continuum spectral indices \hdeltaa\ and \dnfk\ as
indicators of recent star formation, following
\citet{kauffmann03}. Strong \hdelta\ absorption lines arise in
galaxies that experienced a burst of star formation that ended
$\sim$0.1--1 Gyr ago, and large \hdeltaa\ values indicate
strong recent starburst activity
\citep[e.g.,][]{worthey97,kauffmann03c}. \dnfk\ measures the
continuum break around rest-frame 4000 \angstrom , which arises
from a series of metal lines typical of 1--2-Gyr old stars
\citep[e.g.,][]{bruzual83,balogh99,bc03}. \dnfk\ is small for
young stellar populations and large for old, metal-rich
galaxies, and can be used to characterize the
luminosity-weighted mean stellar age \citep{kauffmann03c}.

\begin{figure*}
  \centering
    \includegraphics[width=85mm]{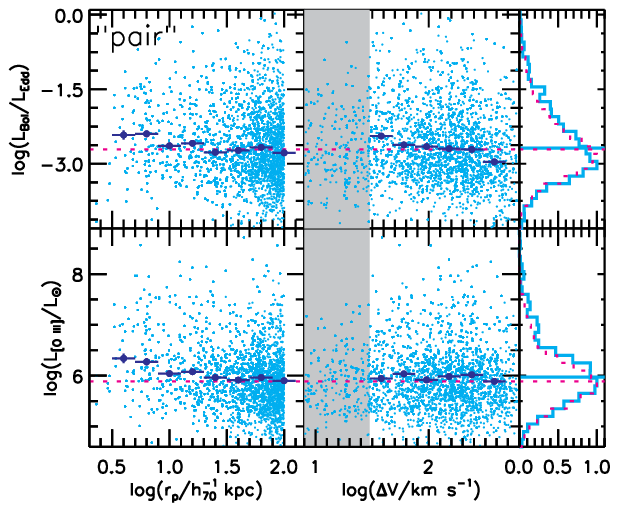}\qquad
    \includegraphics[width=85mm]{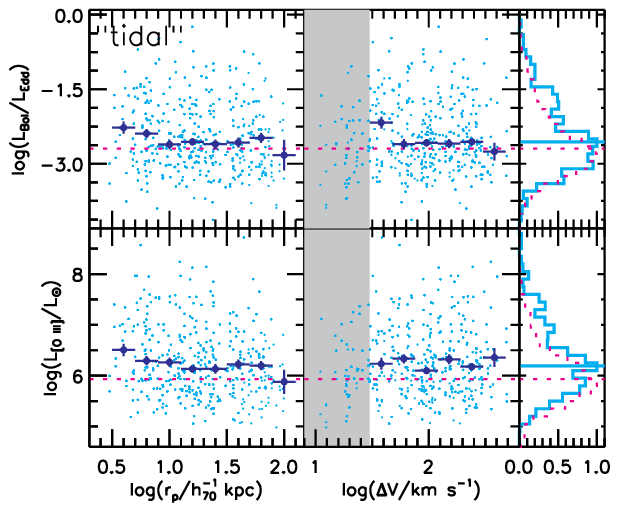}
    \caption{Eddington ratio and \OIIIb\ emission-line luminosity
    (in solar units) as functions of interaction parameters for
    the pair (left block) and tidal (right block) samples.
    Symbols and line types are the same as in Figure \ref{fig:control_at_fixed_r}.}
    \label{fig:agn}
\end{figure*}

Figure \ref{fig:sf} shows \hdeltaa\ and \dnfk\ as functions of
$r_p$ and $\Delta v$ for the pair and tidal samples. Here and
throughout, the error bars denote 1-$\sigma$ uncertainties in
the medians. For AGN pairs with tidal features, both \hdeltaa\
and \dnfk\ are significantly correlated with $r_p$, despite
having a large scatter. Spearman correlation tests show that
the probability of null correlation $P_{null}$ between
\hdeltaa\ (\dnfk ) and $r_p$ is $9\times 10^{-5}$ ($7\times
10^{-6}$). The correlation probabilities quoted are estimated
from the individual data points directly. Pairs with smaller
projected separations on average exhibit stronger recent
starburst activity and younger mean stellar ages. Compared to
the control AGN sample matched in redshift and stellar mass
distributions, AGN pairs with tidal features on average have
larger \hdeltaa\ and \dnfk\ (median values of $1.4\pm0.1$ and
$1.55\pm0.01$ compared to $0.1\pm0.1$ and $1.67\pm0.01$). The
enhancement becomes significant  ($> 3\sigma$ above the
control) for objects with $r_p \lesssim $10--$30$ \hseventy\
kpc (i.e., log$(r_p/$\hseventy\ kpc) $\lesssim 1.0$--$1.5$). In
a sample of $\sim10^5$ star-forming galaxies from SDSS DR4
\citep{SDSSDR4}, \citet{brinchmann04} find that \dnfk\
anti-correlates with specific star formation rate (SSFR)
measured from optical emission lines. Objects with the smallest
separation in our sample ($r_p\sim 5$ \hseventy\ kpc) have
median \dnfk\ $\sim 1.40\pm0.04$, corresponding to log$({\rm
SFR}/M_{\ast})/({\rm yr}^{-1})\sim -10.1$ according to the
calibration of \citet{brinchmann04}. This is $\sim0.9\pm0.2$
dex higher\footnote{This enhancement is reduced to
$\sim0.7\pm0.2$ dex if we adopt the calibration using mean
instead of median values \citep[Figure 11 of][]{brinchmann04}.}
than the SSFRs inferred for the control sample (with median
\dnfk\ $\sim 1.67\pm0.01$, corresponding to log$({\rm
SFR}/M_{\ast})/({\rm yr}^{-1})$ of $-11.0$).

We detect a similar dependence of \hdeltaa\ (\dnfk ) on $r_p$
in the parent population of all AGN pairs (i.e., regardless of
the detection of tidal features). This suggests that the
$r_p$-dependence of \hdeltaa\ (\dnfk ) is robust against biases
from the requirement of tidal features. However, the difference
between the median and overall \hdeltaa\ (\dnfk ) distributions
of the pair and control samples is much smaller than that
between the tidal and control samples, most likely due to
contamination from large-separation AGN pairs that are not yet
interacting. On the other hand, our results seem to suggest
that AGN pairs with tidal features show enhanced recent star
formation at separations as large as $\sim100$ \hseventy\ kpc.
Another possibility for this apparent enhancement at large
separations is that AGN pairs with tidal features by
construction are biased toward interacting systems with recent
or more pronounced star formation because younger stellar
streams have higher surface brightness.

In the tidal sample, we find no significant dependence of
\hdeltaa\ (\dnfk ) on $\Delta v$, after excluding small values
of $\Delta v$ ($<30$ km s$^{-1}$) which are dominated by
redshift measurement uncertainties. Spearman correlation tests
show that the probability of null correlation between \hdeltaa\
(\dnfk ) and $\Delta v$ is 0.72 (0.25). In the pair sample, we
detect a weak dependence ($P_{null}\sim 10^{-3}$) of \hdeltaa\
(\dnfk ) on $\Delta v$, such that pairs with smaller velocity
offsets exhibit stronger recent starburst activity and younger
mean stellar ages.

A caveat is that \hdeltaa\ and \dnfk\ are measured based on
SDSS fiber spectra, which are subject to aperture bias due to
the radial variation of star formation within galaxies
\citep[e.g.,][]{brinchmann04}.  At the smallest and largest
redshifts of our sample ($z=0.02$ and 0.16), the 3$''$ fiber
corresponds to 1.2 kpc and 8.3 kpc, respectively, in the
assumed cosmology. However, the $r_p$ distribution at any fixed
redshift after correction for spectroscopic incompleteness does
not depend significantly on redshift over the range considered,
and therefore our results on the $r_p$ dependence are robust
against aperture bias. Repeating our analysis using SSFR
estimates corrected for aperture bias \citep{brinchmann04}
provided in the MPA-JHU DR7 data product gives results
consistent with those above\footnote{The correction is made by
applying the SSFR distribution at any given $g-r$ and $r-i$
color to SDSS photometry outside the fiber
\citep{brinchmann04}. We use these estimates for verification
purposes only, and take the directly measured spectral indices
(after subtracting emission lines) as our baseline values
because they are less prone to population-dependent bias
\citep[e.g.,][]{salim07}.}.

We must caution that our results are limited to $r_p \gtrsim 5$
\hseventy\ kpc. Our sample is incomplete at $r_p < 5$
\hseventy\ kpc (see Paper I for details), because pairs with
angular separations smaller than $\sim 2''$ will not be
resolved by the deblending algorithm of SDSS photometry and
therefore will not be included in our sample; at $z\sim 0.16$,
i.e., the highest redshift bin of our sample, this corresponds
to $\sim 5$ \hseventy\ kpc. We estimate that the incompleteness
level due to this resolution limit at $r_p = 5$ \hseventy\ kpc
is $\sim 1$--5\% (due to the highest redshift bin around $z =
0.16$), according to the number of smaller separations pairs
($r_p \sim 1$--5 \hseventy\ kpc) resolved at lower redshift in
our sample assuming that the $r_p$ distribution around $r_p\sim
5$ \hseventy\ kpc does not vary with redshift. This
incompleteness level increases with decreasing $r_p$, and it
reaches $\sim 100$\% at $r_p\sim0.8$ \hseventy\ kpc (i.e., the
physical scale corresponding to $\sim2''$ at $z = 0.02$, the
lower redshift boundary of our sample).

\subsubsection{Black Hole Accretion}\label{subsubsec:agn}

We now address how galaxy tidal interactions affect BH
accretion in the early stage of mergers in AGN pairs, and
compare these effects with that of recent star formation to
constrain their causal relationship. We adopt the \OIIIb\
emission-line luminosity \loiii\ and the inferred Eddington
ratio ($L_{{\rm Bol}}/L_{{\rm Edd}}$) as indicators of BH
accretion activity.  We calculate \loiii\ based on fluxes
provided in the MPA-JHU catalogue from Gaussian fits of
continuum-subtracted emission-line spectra. The forbidden line
\OIIIb\ arises from the narrow-line region\footnote{We have
subtracted the contribution from star formation to \loiii\
based on the position on the BPT diagram \citep{bpt,veilleux87}
which features the diagnostic emission-line ratios \OIIIb
/\hbeta\ and \NIIb /\halpha , following the empirical method of
\citet{kauffmann09}. This approach constructs composite objects
using a linear combination of the spectra of pure star-forming
galaxies and pure AGNs. Templates of pure star-forming galaxies
are made with objects located at the lower edge of the
star-forming sequence, whereas pure AGNs are built using those
located at the farthest ends of the Seyfert and LINER regimes.
Trajectories on the diagnostic diagram are then produced from
the constructed composites \citep{kauffmann09}.}, and its
luminosity is correlated with broad-band continuum luminosities
in unobscured AGNs \citep[e.g.,][]{kauffmann03,reyes08,shen10},
albeit with a significant scatter \citep[e.g., 0.36 dex in the
correlation with $M_{2500}$;][]{reyes08}. \citet{heckman05}
found that \OIIIb\ luminosity correlates well with the 3--20
keV luminosity with a dynamic range in luminosity of 4 dex in a
sample of 47 hard X-ray selected local AGNs. To infer
bolometric AGN luminosity $L_{{\rm Bol}}$, we use \loiii\
uncorrected for dust extinction and the appropriate bolometric
correction factor \citep[3500;][]{heckman05,liu09,shen10}. We
do not correct \loiii\ for dust extinction from the hosts in
view of the uncertainties in the assumed reddening laws, which
rely on over-simplified dust-screen models
\citep[e.g.,][]{reyes08}.
For the Eddington luminosity,
%
%
we adopt a BH mass ($M_{{\rm BH}}$) inferred from the stellar
velocity dispersion ($\sigma_{\ast}$) measured from the SDSS
spectrum, assuming the $M_{{\rm BH}}$--$\sigma_{\ast}$ relation
observed in local inactive galaxies
\citep[e.g.,][]{ferrarese00,gebhardt00} and calibrated by
\citet{tremaine02}. As we will discuss below, this is likely an
overestimate of the true mass for small-separation AGN pairs
due to the effects of tidal interactions on the internal
kinematics of the host stellar populations.

\begin{figure*}
  \centering
    \includegraphics[width=85mm]{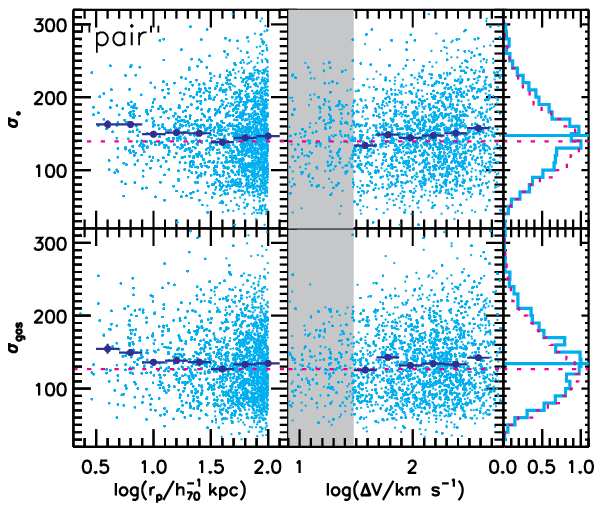}\qquad
    \includegraphics[width=85mm]{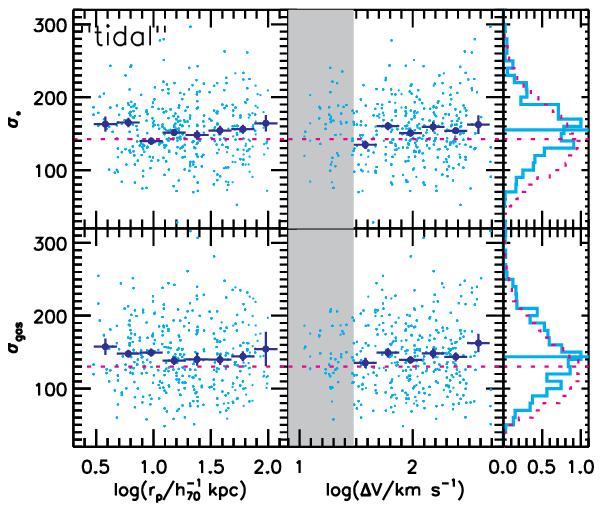}
    \caption{Stellar and gas velocity dispersions as
    functions of interaction parameters for
    the pair (left block) and tidal (right block) samples.
    Symbols and line types are the same as in Figure \ref{fig:control_at_fixed_r}.}
    \label{fig:disp}
\end{figure*}

Figure \ref{fig:agn} shows \OIIIb\ luminosity and Eddington
ratio as functions of $r_p$ and $\Delta v$.  In the tidal
sample, the strength of BH accretion activity increases with
decreasing $r_p$. There is also a similar dependence seen with
$\Delta v$ after excluding small values ($<30$ km s$^{-1}$)
which are dominated by redshift measurement uncertainties.
Spearman correlation tests show that the probability of null
correlation between \loiii\ ($L_{{\rm Bol}}/L_{{\rm Edd}}$) and
$r_p$ is $10^{-3}$ ($2\times 10^{-3}$). These weak correlations
are not as significant as those found between recent star
formation indicators and $r_p$. The enhancement in AGN activity
is significant for objects with $r_p \lesssim 10$--$30$
\hseventy\ kpc (i.e., log$(r_p/$\hseventy\ kpc) $\lesssim
1.0$--$1.5$). Objects with the smallest separation in our
sample ($r_p \sim 5$ \hseventy\ kpc) have median log(\loiii
/$L_{\odot}$) and log($L_{{\rm Bol}}/L_{{\rm Edd}}$) values
$\sim 0.7\pm0.1$ and $\sim0.5\pm0.1$ dex larger than the median
values of the control AGNs.  AGN pairs with tidal features on
average exhibit accretion activity stronger by $\sim$0.2--0.3
dex than control AGNs matched in redshift and stellar mass
distributions.

In the pair sample we detect a similar dependence of \loiii\
($L_{{\rm Bol}}/L_{{\rm Edd}}$) on $r_p$. This verifies that
the $r_p$-dependence result is robust against biases from our
identification of tidal features.  However, the differences in
the median value and distribution of \loiii\ ($L_{{\rm
Bol}}/L_{{\rm Edd}}$) between the pair and control samples are
much smaller than those between the tidal and control samples.
The dependences of \loiii\ and $L_{{\rm Bol}}/L_{{\rm Edd}}$ on
$r_p$ are less prominent than that of recent star formation
shown in Figure \ref{fig:sf}, and therefore can be more
difficult to detect.

Figure \ref{fig:disp} shows that $\sigma_{\ast}$ at small $r_p$
($\sim 10$--30 \hseventy\ kpc) is on average larger than that
of large separation pairs and the median value of control AGNs
by $\sim20\pm5$ km s$^{-1}$ in both the pair and tidal samples.
The stellar mass and effective radius are independent of $r_p$.
Therefore the larger $\sigma_{\ast}$ cannot be due to deeper
potentials; it is most likely the effect of the interactions
with the transformation of orbital into internal energy. As a
result, the inferred BH mass is likely overestimated by
$\sim0.2$ dex on average, and by extension, the enhancement of
$L_{{\rm Bol}}/L_{{\rm Edd}}$ with decreasing $r_p$ is likely
underestimated by $\sim0.2$ dex in AGN pairs with the smallest
separations in our samples.  We see a similar enhancement in
gas velocity dispersion (measured from Gaussian fits to the
forbidden emission lines) as well (Figure \ref{fig:disp}).

Because the redshift distribution corrected for spectroscopic
incompleteness at any fixed $r_p$ does not depend significantly
on $r_p$ (Paper I), our results involving luminosities are
robust against redshift-induced biases. Repeating our analysis
using \OIIIb\ equivalent width (EW) as an indicator of BH
accretion yields consistent results.  Another caveat is that
the \OIIIb\ fluxes are based on SDSS fiber spectra and could
therefore be subject to aperture bias (see the discussion at
the end of \S \ref{subsubsec:sf}). The typical size of the
narrow-line regions of lower-luminosity AGNs is several hundred
parsecs, which is smaller than the aperture size of objects
with the lowest redshift in the sample, and therefore the bulk
of narrow-line region \OIIIb\ emission should be included in
the fiber spectra of all objects in the sample. In addition, we
have verified that the distribution of separations for $5 < r_p
< 100$ \hseventy\ kpc at each redshift bin (after correction
for spectroscopic incompleteness) does not vary significantly
with redshift (over the range that we consider, i.e., $0.02 < z
< 0.16$). Therefore, the correlations of the galaxy properties
with $r_p$ are not affected by redshift dependent biases.

\begin{figure*}
  \centering
    \includegraphics[width=80mm]{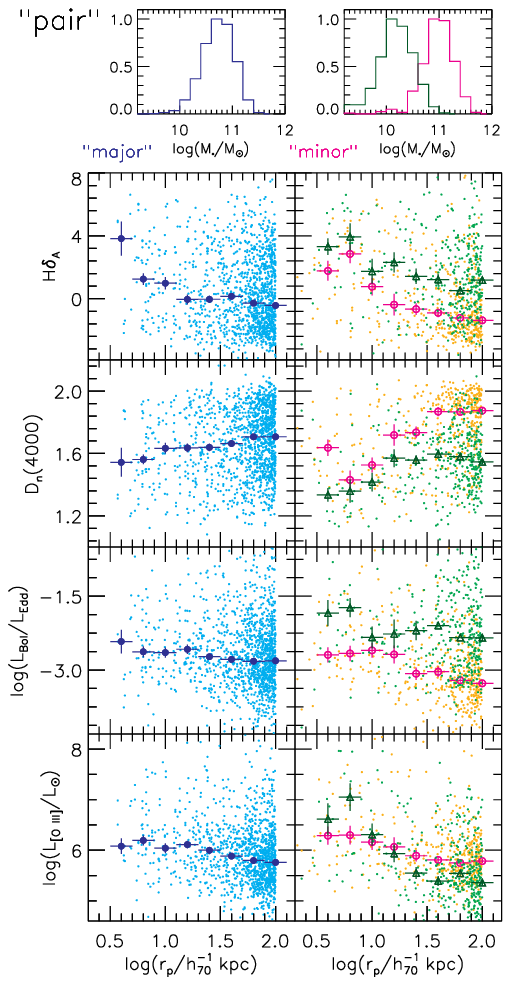}\qquad
    \includegraphics[width=80mm]{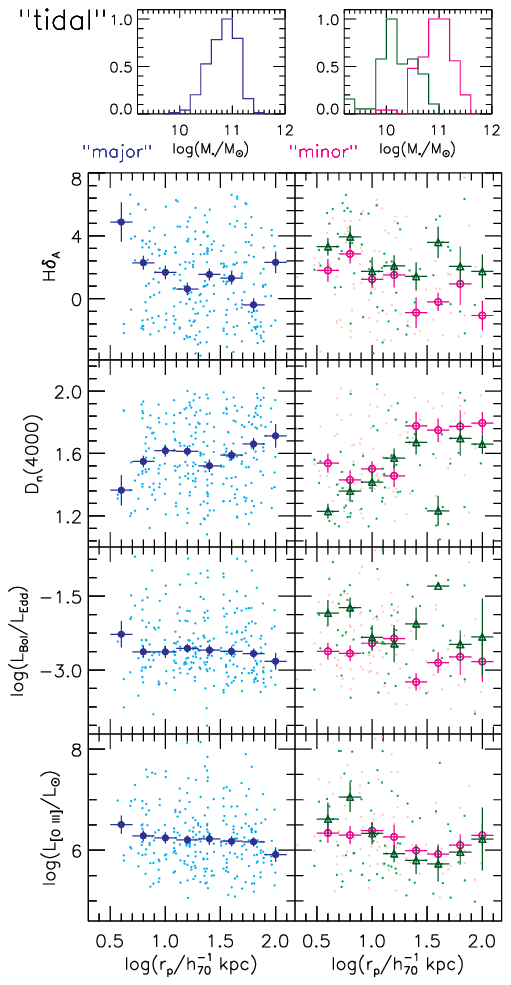}
    \caption{The effect of host-galaxy mass ratio on tidally
    enhanced star formation and AGN activity.
    The left (right) block is for the pair (tidal) sample.
    In each block, the left column is for major mergers
    with mass ratios smaller than 3,
    whereas the right column is for minor mergers with mass ratios larger than 3.
    For minor mergers,
    we show the primary (open circles in magenta) and secondary
    (open triangles in green) galaxies separately.
    Small dots denote individual objects, whereas large symbols
    represent median values at each $r_p$ bin.
    Error bars indicate uncertainties of the median values.
    The panels in the first row display
    stellar mass distributions.}
    \label{fig:sf:massratio}
\end{figure*}

\subsection{Dependence on Host-Galaxy Mass and Structure}\label{subsec:progenitor}

We now address how host-galaxy internal properties regulate the
effects of tidal interactions on recent star formation and AGN
activity in AGN pairs.  We examine the role of mass ratio in \S
\ref{subsubsec:massratio} and that of galaxy structure (as
indicated by concentration index) in \S \ref{subsubsec:bulge}.
We focus on the dependence of recent star formation and BH
accretion as a function of $r_p$.

\subsubsection{Effect of Host-Galaxy Mass Ratio: Major versus Minor
Mergers}\label{subsubsec:massratio}

Mergers and close interactions (hereafter mergers for short) of
galaxy pairs with comparable masses (so called ``major
mergers'') as a driver for star formation and AGN activity have
been the topic of extensive theoretical and observational
research \citep[e.g., see reviews by][]{barnes92,struck06}. It
is generally accepted that major mergers of gas-rich spirals
are responsible for the most luminous starburst galaxies,
including the nearby ultra-luminous infrared galaxies
\citep{sanders88,hernquist89,mihos96} and sub-millimeter
galaxies at higher redshifts
\citep{blain02,conselice03b,tacconi08}. Minor mergers are more
frequent than major mergers. At $z<1$ there is evidence that
minor mergers may play a dominant role in inducing moderate
star formation in early-type galaxies \citep{kaviraj10} and in
driving the structural evolution of intermediate-mass galaxies
\citep{lopez10}. It is still debated whether major or minor
mergers or internal instabilities are the primary triggers for
various AGN populations \citep[for a review see][]{jogee06}.

We divide the our AGN pair samples into major and minor
mergers, i.e., with stellar mass ratios smaller and larger than
$3:1$, respectively. The stellar mass ratios of our sample span
the range $1 \leq M_{\ast,1}/M_{\ast,2} \lesssim 10$, where
$M_{\ast,1}$ denotes the stellar mass of the primary and
$M_{\ast,2}$ is that of the secondary galaxy. 38\% of our AGN
pairs have mass ratios smaller than 3. The sample is incomplete
for pairs with mass ratios larger than $\sim$10, because of the
limited dynamical range of the parent spectroscopic galaxy
sample.

\begin{figure*}
  \centering
    \includegraphics[width=80mm]{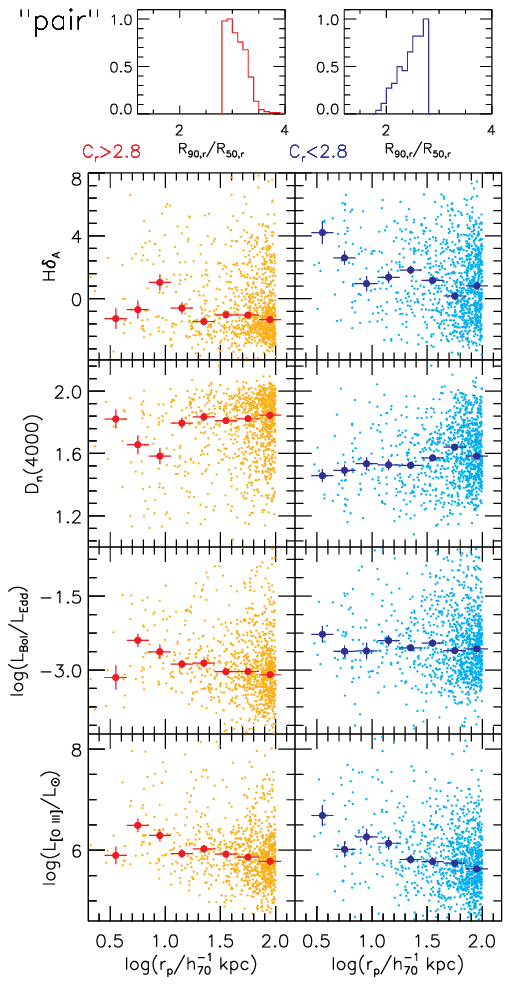}\qquad
    \includegraphics[width=80mm]{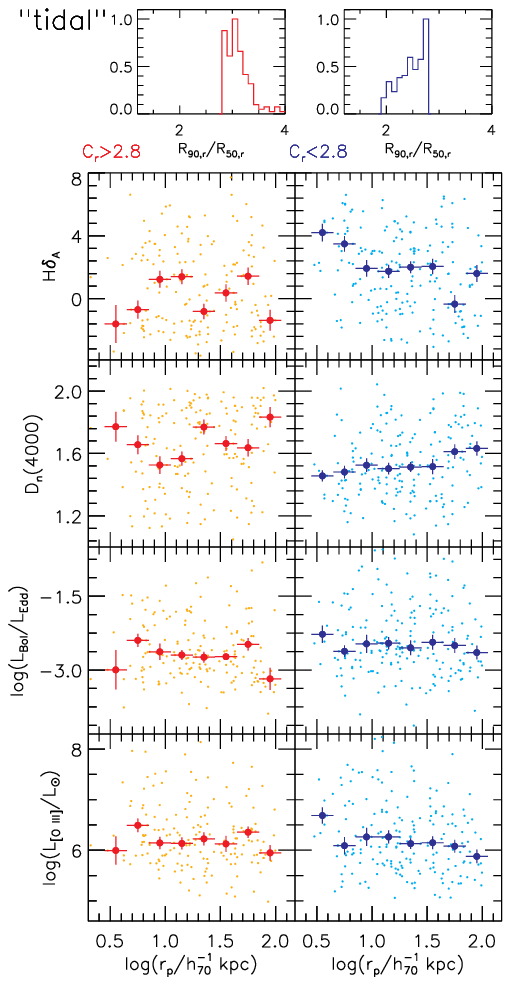}
    \caption{The effect of host-galaxy structure on tidally enhanced star formation and
    AGN
    activity.
    The left (right) block is for the pair (tidal) sample.
    In each block, the left column is for galaxies with $r$-band
    concentration index $C_r > 2.8$ (corresponding roughly to ellipticals)
    whereas the right column is for those with $C_r < 2.8$ (spirals).
    Small dots denote individual objects, whereas large symbols
    represent median values at each $r_p$ bin.
    Error bars indicate uncertainties of the median values.
    The panels in the first row show the $C_r$ distributions.}
    \label{fig:sf:bulge}
\end{figure*}

Figure \ref{fig:sf:massratio} shows the recent star formation
and AGN indicators as functions of $r_p$ for major and minor
mergers for the pair and tidal samples. In the pair sample,
galaxies in major mergers have median stellar masses of $\sim
10^{10.7} M_{\odot}$, whereas the primary and secondary of
minor mergers have median stellar masses of $\sim 10^{10.2}
M_{\odot}$ and $\sim 10^{11.0} M_{\odot}$, respectively.  We
observe a dependence of host recent star formation and AGN
activity on $r_p$ in both major and minor mergers. In minor
mergers, we detect and enhancement of recent star formation and
AGN activity indicators at small $r_p$ in both the primary and
secondary galaxies.  AGN pairs with separations of $\sim5$
\hseventy\ kpc have median SSFRs that are larger than those
with separations of $r_p\sim100$ \hseventy\ kpc (approaching
the case of control samples of single AGNs) by $\sim0.8$,
$\sim0.6$, and $\sim0.7$ dex (with typical uncertainties of 0.2
dex) in major, minor-primary, and minor-secondary galaxies
respectively, based on the SSFR-\dnfk\ calibration of
\citet{brinchmann04}.  The enhancement level of median \loiii\
($L_{{\rm Bol}}/L_{{\rm Edd}}$) in AGN pairs from $r_p\sim100$
to $\sim5$ \hseventy\ kpc is 0.4, 0.5, and 1.2 dex (0.4, 0.6,
and 0.7 dex) for major, minor-primary, and minor-secondary
galaxies, respectively, with typical uncertainties of 0.2, 0.4,
and 0.3 dex.  The most notable difference between the secondary
and primary in minor mergers is that the enhancement level of
median \loiii\ from $r_p\sim100$ to $\sim5$ \hseventy\ kpc is
significantly higher in the secondary than in the primary
galaxy.  The difference between the secondary and primary in
the enhancement of $L_{{\rm Bol}}/L_{{\rm Edd}}$ is less
significant, although we find that the enhancement of
$\sigma_{\ast}$ is larger in the secondary than in the primary
which may cause $L_{{\rm Bol}}/L_{{\rm Edd}}$ to be more
severely underestimated (see the discussion at the end of \S
\ref{subsubsec:agn}). In the tidal sample, we find similar
results to those in the pair sample, except that there is a
significantly larger enhancement of \dnfk\ in the secondary
than in the primary of minor mergers. We discuss the
implications of these results in \S \ref{sec:discuss}.

\begin{figure}
  \centering
    \includegraphics[width=85mm]{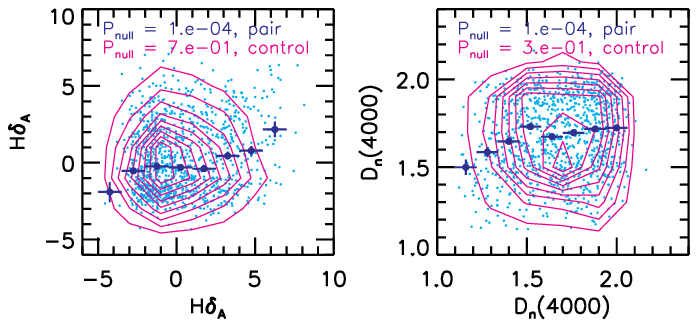}
    \includegraphics[width=85mm]{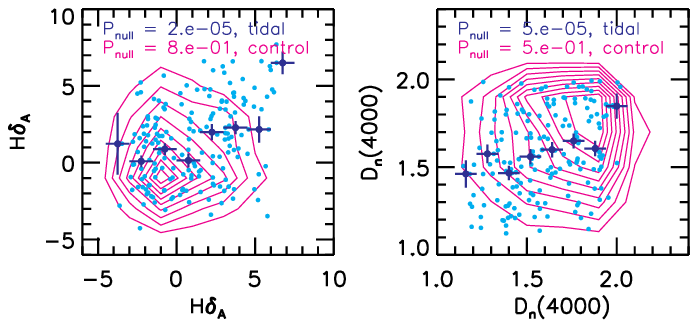}
    \caption{Correlation of host-galaxy recent star formation
    as indicated by the spectral indices \hdeltaa\ and \dnfk\ between the interacting
    components in each AGN pair.
    The top (bottom) row is for the pair (tidal) sample.
    The $x$ axis is for one component (randomly pulled from the two in a pair)
    and the $y$ axis is for the other component.
    Cyan dots denote individual objects, whereas blue filled circles
    represent the median value of objects binned in each interval.
    Error bars indicate uncertainties of the median.
    Contours in magenta indicate ten evenly spaced
    number densities of control samples of AGNs each matched in redshift
    and in stellar mass with one of the two AGNs in a pair.
    Displayed on each panel are the Spearman
    probabilities of null correlation for the control (magenta)
    and the AGN pair/tidal samples (blue).}
    \label{fig:corr:sf}
\end{figure}

\begin{figure}
  \centering
    \includegraphics[width=41mm]{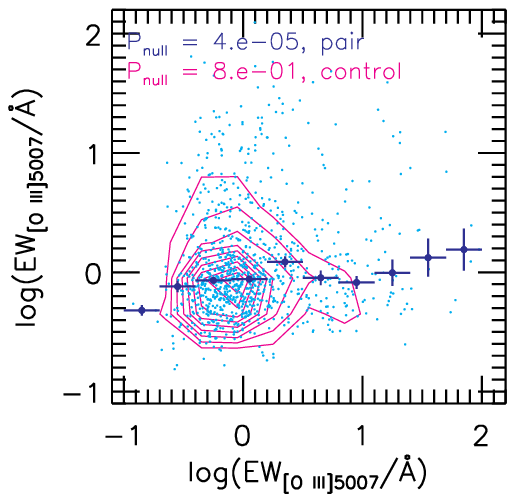}\quad
    \includegraphics[width=41mm]{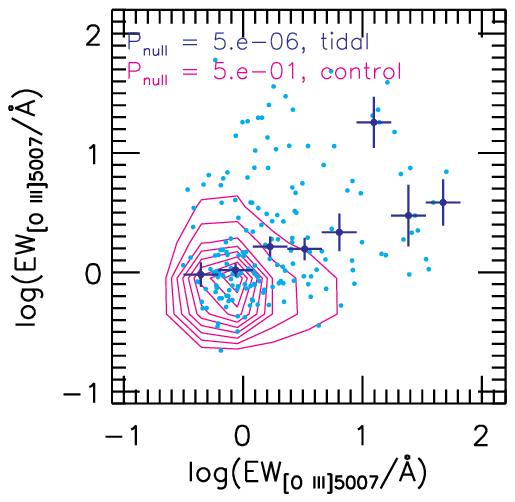}
    \caption{As in Figure \ref{fig:corr:sf}, but for
    the correlation of the rest-frame \OIIIb\ EW
    (an indicator of BH accretion activity) between
    the interacting components in each AGN pair.
    The left (right) panel is for the pair (tidal) sample.}
    \label{fig:corr:agn}
\end{figure}

\subsubsection{Effect of Host-Galaxy Structure: Bulge versus Disk
Mergers}\label{subsubsec:bulge}

The distribution of mass within galaxies regulates when
gravitational instabilities are induced in galaxy tidal
interactions. Simulations of major mergers of disk galaxies
show that a dynamically hot bulge component stabilizes a disk
against bar instability, thus delaying merger-induced activity
\citep{mihos96}. We address whether such an effect is present
in AGN pairs. We use the $r$-band concentration index, $C_r$,
as a measure of mass distribution.  \citet{strateva01} have
shown that $C_r$ correlates with galaxy morphological type
albeit with a large scatter \citep[see also][]{shimasaku01};
elliptical galaxies have larger $C_r$ than do spirals. $C \sim
5.5$ for a pure de Vaucouleurs elliptical whereas $C \sim 2.3$
for a pure exponential disk. In particular, \citet{strateva01}
have found that a value of $C_r = 2.83$ divides early- and
late-type galaxies with a reliability of 83\% for each and with
completeness of 70\% and 80\%, respectively.  As shown in
Figure \ref{fig:sf:bulge}, the $C_r$ distribution of our AGN
pair samples is close to being a Gaussian with a median value
of $C_r = 2.8$, and lies in the intermediate region between
early- and late-type galaxies. We divide the AGN pair galaxies
into those with $C_r > 2.8$ and those with $C_r < 2.8$ (each
component in a pair is treated separately).

We first examine how host-galaxy structure regulates the
effects of tidal interactions on host star formation in AGN
pairs. Figure \ref{fig:sf:bulge} shows the dependence of recent
star formation and AGN activity indicators on $r_p$ for early-
and late-type galaxies in the pair and tidal samples,
respectively. In both samples we find a clear dependence of
host star formation indicators on $r_p$ in AGNs with late-type
host galaxies. From $r_p\sim100$ to $\sim5$ \hseventy\ kpc, the
median SSFR values are enhanced by $\sim0.4$ dex in AGNs with
late-type hosts, according to the SSFR-\dnfk\ calibration of
\citet{brinchmann04}.  In AGNs with early-type hosts, however,
the trend is much less prominent. In the pair sample, neither
\hdeltaa\ and \dnfk\ show much dependence on $r_p$ until $r_p
\lesssim 10$ \hseventy\ kpc, whereas the dependence becomes
prominent starting at $r_p \lesssim 30$ \hseventy\ kpc in
late-type hosts. In the tidal sample, there is little
dependence of \hdeltaa\ and \dnfk\ on $r_p$. Our comparison
between the early- and late-type hosts of AGN pairs suggest
that tidally enhanced star formation happens earlier and more
prominent in late- than in early-type hosts of AGN pairs.

We now assess how host-galaxy structure regulates the effects
of tidal interactions on BH accretion in AGN pairs.  As shown
in Figure \ref{fig:sf:bulge}, the median \loiii\ ($L_{{\rm
Bol}}/L_{{\rm Edd}}$) value increases with decreasing $r_p$ in
AGNs with both early- and late-type host galaxies in both the
pair and tidal samples.  In the pair sample, the median \loiii\
value increases from $r_p\sim100$ to $\sim5$ \hseventy\ kpc by
0.7$\pm$0.1 dex (1.0$\pm$0.1 dex) of AGNs with early-type
(late-type) hosts, respectively (here we neglect the smallest
$r_p$ bin in early-type hosts to focus on the overall trend).
The increase of the median $L_{{\rm Bol}}/L_{{\rm Edd}}$ value
seems less prominent: 0.7$\pm$0.1 dex (0.4$\pm$0.1 dex) for
AGNs in early-type (late-type) hosts. However, the enhancement
of $L_{{\rm Bol}}/L_{{\rm Edd}}$ could be underestimated
because $\sigma_{\ast}$, and by extension $M_{{\rm BH}}$, is
likely overestimated for AGN pairs with small separations
(Figure \ref{fig:disp}); late-type hosts show more
$\sigma_{\ast}$ enhancement. We detect similar trends in the
tidal sample. The different levels of \loiii\ median
enhancement in AGNs with early- and late-type hosts suggest
that AGNs with late-type hosts are more subject to tidally
enhanced BH accretion than those with early-type hosts.

We have verified that the trend we see in the early and
late-type hosts of AGN pairs in indeed due to the difference in
their concentration rather than a mass selection effect. First,
we compared the stellar masses of the $C_r > 2.8$ and the $C_r
< 2.8$ samples (here we quote the numbers for the pair sample)
and found that they defer by only 0.2 dex on average; their
stellar mass distributions are consistent with being Gaussian
distributions, with the $C_r > 2.8$ sample having a median mass
of 10.81 (in units of log$(M_{\ast}/M_{\odot})$) and a standard
deviation of 0.34, and the $C_r < 2.8$ sample having a median
stellar mass of 10.62 and a standard deviation of 0.43. This
0.2 dex difference is much smaller than that between the
primaries and secondaries ($\sim1$ dex) in minor merger pairs
(\S \ref{subsubsec:massratio}). While the $C_r > 2.8$ sample
has on average larger stellar masses than the $C_r < 2.8$
sample (as expected due to the mass-concentration correlation),
the stellar mass difference seems to be too small to be the
dominant factor of the trend we see. Second, to test whether
the effect of this small mass difference is indeed minor, we
redid our analysis using subsets of the $C_r > 2.8$ and the
$C_r < 2.8$ samples matched in stellar mass distribution. The
trend was still present, verifying that it was indeed due to
concentration effect rather than a mass selection effect.

\subsection{Correlations between the Interacting Components in an AGN
Pair}\label{subsec:correlation}

In this section we examine the correlations between the two
interacting components in each AGN pair. The statistical sample
of AGN pairs presented in Paper I allows us to address whether
there is a correlation in the strength of BH accretion between
the interacting components. We show in Figure \ref{fig:corr:sf}
the correlations in indicators of recent star formation between
the interacting components for the pair and tidal samples.  We
plot \hdeltaa\ (\dnfk ) of one component (randomly pulled from
the two) against that of the other component in an interacting
pair. Also shown for comparison are density contours from
control AGNs. We draw control samples of AGNs for both
components, each of which is matched in redshift and stellar
mass distributions to one component of the AGN pairs. The two
control samples are then randomly ``paired'' with the
requirement that each control ``pair'' has a LOS velocity
offset smaller than 600 km s$^{-1}$, although our results
remain the same without this constraint. As shown in Figure
\ref{fig:corr:sf}, there is no correlation between the two
components in the control sample; Spearman null probabilities
are close to 1.  We have run 1000 Monte Carlo simulations for
the control sample and verified that the Spearman correlation
probabilities shown in Figure \ref{fig:corr:sf} are typical.
For the pair and tidal samples, on the other hand, we detect a
statistically significant correlation between the recent star
formation indicators of the interacting galaxies. Spearman
tests show the probability of null correlation is $1\times
10^{-4}$ ($1\times 10^{-4}$) in \hdeltaa\ (\dnfk ) for the pair
sample; the null probability is even smaller for the tidal
sample, $2\times 10^{-5}$ ($5\times 10^{-5}$) in \hdeltaa\
(\dnfk ). There is a substantial scatter in the correlations
though.  The correlations seen in Figure \ref{fig:corr:sf} for
AGN pairs are reminiscent of the color correlation seen between
inactive galaxy pairs \citep[the Holmberg effect; e.g.,
][]{holmberg58,madore86,kennicutt87}, which is generally
accepted as evidence for merger-induced star formation
\citep[e.g.,][]{laurikainen89}.

\begin{figure}
  \centering
    \includegraphics[width=85mm]{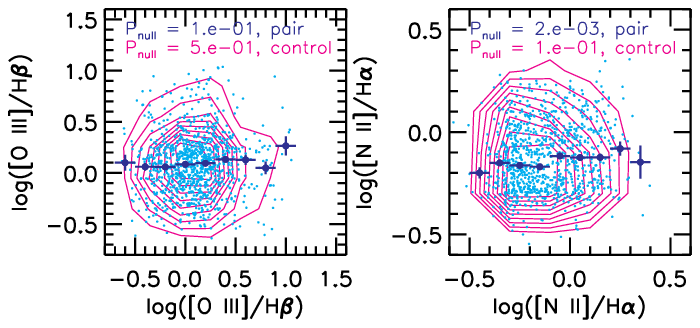}\qquad
    \includegraphics[width=85mm]{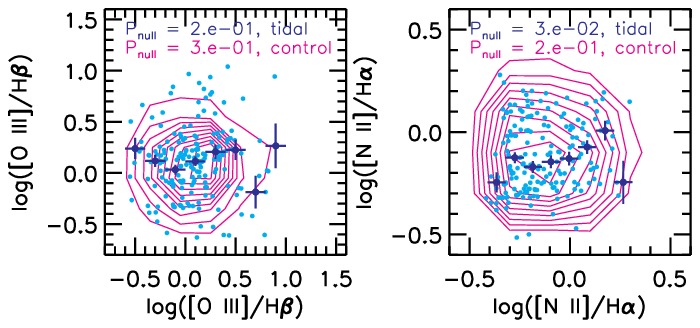}
    \caption{As in Figure \ref{fig:corr:sf}, but for
    the correlation of the diagnostic emission-line ratios
    between the interacting components in each AGN pair.
    The top (bottom) row is for the pair (tidal) sample.}
    \label{fig:corr:bpt}
\end{figure}

We now address whether there is an effect in BH accretion
activity similar to that observed in recent star formation. To
avoid selection bias due to the common distance moduli in a
pair in a flux-limited sample, here we adopt the rest-frame
\OIIIb\ equivalent width (EW) as an indicator of BH accretion,
instead of \OIIIb\ luminosity or Eddington ratio. Figure
\ref{fig:corr:agn} shows the \OIIIb\ EW of one component
(randomly pulled from each AGN pair) as a function of that of
the other component in the pair for the pair and tidal samples.
In the pair sample, we detect a statistically significant
correlation in the \OIIIb\ EW, although the scatter is
significant ($\sim 1$ dex); the Spearman null probability is
$4\times 10^{-5}$. The correlation in the tidal sample is even
stronger, with a null probability of $5\times 10^{-6}$. We find
no correlation between \OIIIb\ EWs of the two components in the
control samples; the Spearman null probabilities are close to
1. The correlation of AGN activity between the two components
in an AGN pair suggests that there is tidally enhanced AGN
activity in both components. Alternatively, a single active
SMBH ionizing both galaxies could also produce such a
correlation, at least for the subset of AGN pairs with
separations smaller than galaxy sizes.  The latter scenario is
unlikely, however, as we find no significant
correlation\footnote{There is a tentative correlation of \NII
/\halpha\ in both the pair and tidal samples, which could be
due to the tidal effect on the gas-phase metallicity observed
in inactive galaxy pairs \citep[e.g.,][]{ellison08}.} in the
diagnostic emission-line ratios \OIII /\hbeta\ \footnote{The [O
\OIII /\hbeta\ ratios have not been corrected for contributions
from star formation. The results show no significant change
when \OIII\ is corrected for star formation.} between the
interacting components as shown in Figure \ref{fig:corr:bpt},
suggesting that the two components are not powered by the same
central engine.

\subsection{Correlations between Recent Star Formation and Black Hole
Accretion}\label{subsec:sfagn}

AGN luminosity is correlated with \hdeltaa\ and \dnfk\ in the
host galaxies of $z\sim 0.1$ AGNs, in the sense that the hosts
of higher-luminosity AGNs have younger stellar populations and
higher starburst fractions in the last 0.1-1 Gyr
\citep[e.g.,][]{kauffmann03,netzer09,liu10c}. Here we address
whether AGN luminosity is correlated with recent star formation
in hosts in AGN pairs, and how such a correlation, if present,
compares to that in ordinary AGNs.

\begin{figure*}
  \centering
    \includegraphics[width=85mm]{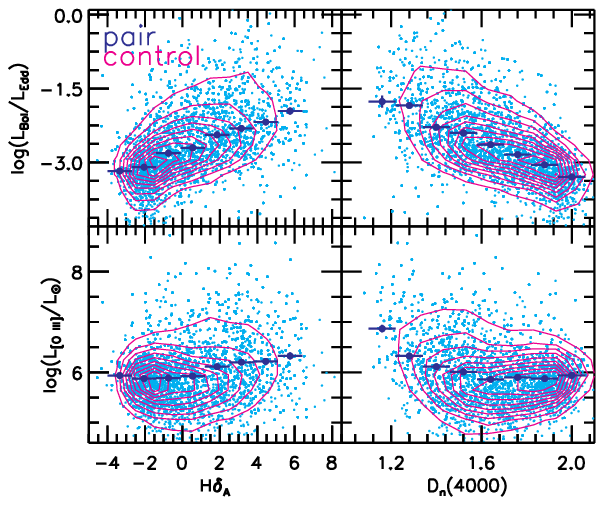}\qquad
    \includegraphics[width=85mm]{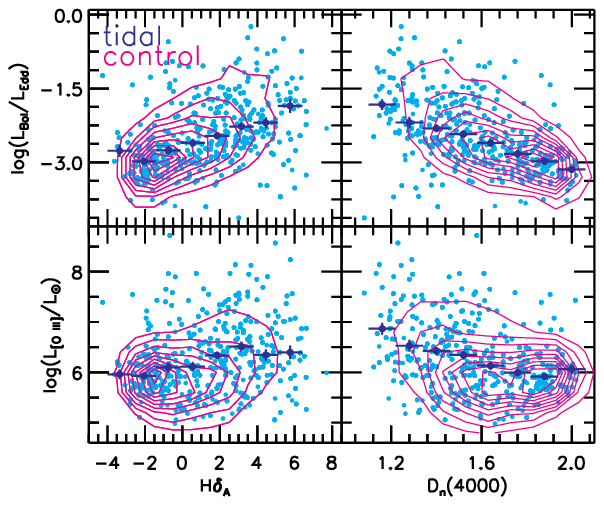}
    \caption{Correlation between BH accretion
    (in terms of Eddington ratio and \OIIIb\ emission-line luminosity) and
    recent star formation (as indicated by the spectral
    indices \hdeltaa\ and \dnfk ) in AGN pairs.
    The left (right) block is for the pair (tidal) sample.
    Cyan dots denote individual objects in AGN pairs, whereas
    blue filled circles represent the median value of objects
    binned in each interval.
    Error bars indicate uncertainties of the median.
    Contours in magenta indicate control samples of single
    AGNs matched in redshift and in stellar mass.
    }
    \label{fig:sfagncorr}
\end{figure*}

Figure \ref{fig:sfagncorr} displays \loiii\ and $L_{{\rm
Bol}}/L_{{\rm Edd}}$ as a function of \hdeltaa\ and \dnfk\ for
the pair and tidal samples. Also shown for comparison are
contours from control AGN samples matched in redshift and
stellar mass distributions. There is a correlation between
\loiii\ and \hdeltaa\ (\dnfk ) when \hdeltaa $\gtrsim 2$ (\dnfk
$\lesssim 1.6$) for the pair sample and when \hdeltaa $\gtrsim
-2$ (\dnfk $\lesssim 1.9$) for the tidal sample, respectively.
$L_{{\rm Bol}}/L_{{\rm Edd}}$ is also correlated with \hdeltaa\
and \dnfk\ for both samples, although at least part of the
correlation is driven by the mass dependence of both
$\sigma_{\ast}$ (and by extension $L_{{\rm Edd}}$) and
\hdeltaa\ (\dnfk ). The relation between AGN luminosity and
\hdeltaa\ (\dnfk ) of the pair sample is almost identical to
that of the control AGN sample. While the tidal sample occupies
a similar scaling relation between AGN luminosity and \hdeltaa\
(\dnfk ) to that of the control sample, it is skewed towards
higher AGN luminosities and larger starburst fractions (younger
mean stellar ages).

\section{Discussion}\label{sec:discuss}

We discuss implications of our results on tidally enhanced star
formation and BH accretion. We compare our results with
previous observations of inactive galaxy pairs and of single
AGNs in galaxy pairs in \S \ref{subsec:compare}, and to
predictions from galaxy merger simulations in \S
\ref{subsec:comparesim}.

\subsection{Comparison with Observations of Galaxy Interactions}\label{subsec:compare}

\subsubsection{Tidally Enhanced Star Formation}

We have detected a correlation between \hdeltaa\ (and \dnfk )
and projected separation $r_p$ in AGN pairs; systems with
smaller separations have larger \hdeltaa\ and smaller \dnfk\
indicative of higher SSFRs, with the enhancement becoming
significant for $r_p \lesssim 10$--$30$ \hseventy\ kpc (Figure
\ref{fig:sf}).  We also find a weak correlation between
\hdeltaa\ (and \dnfk ) and $\Delta v$ in both our tidal and
pair samples, after excluding small values of $\Delta v$ which
are dominated by redshift measurement errors.  These results
are in broad agreement with previous findings based on
statistical samples of inactive galaxy pairs
\citep{barton00,lambas03,alonso04,nikolic04,ellison08,li08a,darg10b}.
For example, in a volume-limited sample of 12,492 main galaxies
from the SDSS DR1 \citep{SDSSDR1} selected to have photometric
companions within 300 kpc, \citet{nikolic04} used SFRs inferred
from \halpha\ luminosities to find that the mean SSFR is
significantly enhanced for $r_p < 30$ \hseventy\ kpc.  These
authors also find a weak anti-correlation between SSFR and
recessional velocity difference $\Delta v$ \citep[see
also][]{lambas03}. Similarly, \citet{ellison08} found an
enhancement in the star formation rates of galaxy pairs at $r_p
<$ 30--40 \hseventy\ kpc in a study of 1716 emission-line
galaxies selected from the SDSS DR4 \citep{SDSSDR4} with
companions within $\Delta v < 500$ km s$^{-1}$, $r_p< 80$
\hseventy\ kpc, and stellar mass ratios smaller than 10.

The average SSFR enhancement level we observe in AGN pairs is
$0.7$--$0.9\pm0.2$ dex from $r_p \gtrsim 100$ to $\sim 5$
\hseventy\ kpc. This is comparable to some previous estimates
based on inactive galaxy pairs \citep[e.g., factors of 1.5--4
from $r_p> 100$ to $r_p = 20$ kpc;][]{li08a}, but seems to be
larger than others taken at face value (e.g.,
\citealt{alonso04} and \citealt{darg10b} found an enhancement
of a factor of 2). The variation in the observed enhancement
levels may be due to differences in the galaxy samples studied.
For example, the \citet{alonso04} results are based on galaxy
pairs selected in galaxy groups or clusters with virial masses
in the rage of $10^{13}$--$10^{15} M_{\odot}$ and therefore may
be biased towards early-type galaxies. Indeed, we find
tentative evidence that the star formation enhancement is more
prominent and occurs at larger separations in late-type than in
early-type galaxies (Figure \ref{fig:sf:bulge}).  The variation
in the observed enhancement levels may also be due to
differences in the interaction stages examined. For example,
the factor of 2 difference inferred by \citet{darg10b} refers
to the average difference between their merger sample
(including various separations hence interaction stages) and
the control sample, which may be much smaller than the
difference observed for small separation pairs. Other factors
that may affect the observed level of enhancement include how
control samples are defined and the contamination from closely
separated galaxy pairs that are not tidally interacting.

We find that the enhancement level of recent star formation in
AGN pairs with host stellar mass ratios
$M_{\ast,1}/M_{\ast,2}>3$ (minor mergers or interactions) is
comparable to (if not slightly smaller than) that in those with
stellar mass ratios $M_{\ast,1}/M_{\ast,2}<3$ (major mergers or
interactions), at least when the two galaxies are separated by
more than $r_p \sim 5$ \hseventy\ kpc. In addition, the
relative enhancement level is similar in both the primaries and
the secondaries in AGN pairs with stellar mass ratios
$M_{\ast,1}/M_{\ast,2}>3$.  Similarly, \citet{ellison08}
detected SFR enhancement in their sample of inactive galaxy
pairs drawn from the SDSS whose stellar masses are within a
factor of 10 of each other. On the other hand, based on 167
galaxy pairs drawn from the CfA2 Redshift Survey
\citep{huchra90,huchra95}, \citet{woods06} suggest that
galaxies in major interactions are more likely to show enhanced
star formation activity than those in minor interactions with
absolute magnitude differences larger than 2. In a sample of
1204 galaxies in spectroscopic pairs and compact groups
selected from the SDSS DR5 \citep{SDSSDR5}, \citet{woods07}
find that in minor mergers the secondary, but not the primary
shows tidally triggered star formation.  On the other hand, in
a study of 1258 spectroscopic galaxy pairs drawn from the 2dF
survey \citep{colless01}, \citet{lambas03} find that in minor
mergers the bright components are more likely to show
tidally-enhanced star formation than the faint components. We
speculate that these contradictory results could be due to
difference in the galaxy samples studied, variation in their
sample sizes, and criteria to define major and minor mergers.
For example, we adopt stellar mass ratio in the current study
\citep[see also][]{ellison08}, whereas \citet{lambas03},
\citet{woods06}, and \citet{woods07} employed the ratio of
galaxy luminosities in their studies.

We caution that our conclusion, that recent star formation is
enhanced in AGN pairs during the early merger stages, may not
hold for all types of galaxies, given the inherent limitations
of our sample selection. Low-redshift optical AGNs reside in
galaxies with stellar mass, galaxy structure, and star
formation properties intermediate between young disk-dominated
and old bulge-dominated galaxies
\citep[e.g.,][]{kauffmann03,kauffmann03b,heckman04}. Whether
tidal encounters enhance star formation in galaxies with all
ranges of galaxy mass and structure is beyond the scope of this
paper, although there is mounting evidence that the specific
levels of star formation enhancement depend on interaction
and/or progenitor properties
\citep[e.g.,][]{bushouse86,bushouse88,bergvall03,brosch04,woods06,rogers09}.

\subsubsection{Tidally Enhanced Black Hole Accretion}

Previous statistical studies of galaxy pairs found evidence for
a higher fraction of AGNs in paired than in isolated galaxies
\citep{kennicutt87,alonso07,woods07,rogers09,ellison11} whereas
others detected no significant difference
\citep{barton00,ellison08,li08b,darg10b}. We have found a
correlation between \OIIIb\ luminosity (and Eddington ratio)
and projected separation in AGN pairs; systems with smaller
separations host more powerful AGNs and the enhancement becomes
significant for $r_p \lesssim 10$--$30$ \hseventy\ kpc (Figure
\ref{fig:agn}), although the enhancement is only moderate,
which is $\sim0.2$--0.3 dex on average in luminosity/Eddington
ratio relative to control samples and rises to $\sim0.5$--0.7
dex for the closest pairs in our sample ($r_p\sim5$ \hseventy\
kpc). We find that the {\it relative} enhancement in AGN
luminosity from $r_p \gtrsim 100$ to $\sim 5$ \hseventy\ kpc is
higher in the secondary than in the primary in AGN pairs with
stellar mass ratios $M_{\ast,1}/M_{\ast,2}>3$. Such a
difference may be driven by the fact that less massive galaxies
on average have higher gas fractions, or that they are more
prone to gravitational instabilities. The offset Seyfert 2
nucleus in the minor merger system NGC 3341 discovered by
\citet{barth08} may be a manifestation of the same effect,
where the BH in the secondary is experiencing much more
accretion than that in the primary.

\begin{figure}
  \centering
    \includegraphics[width=85mm]{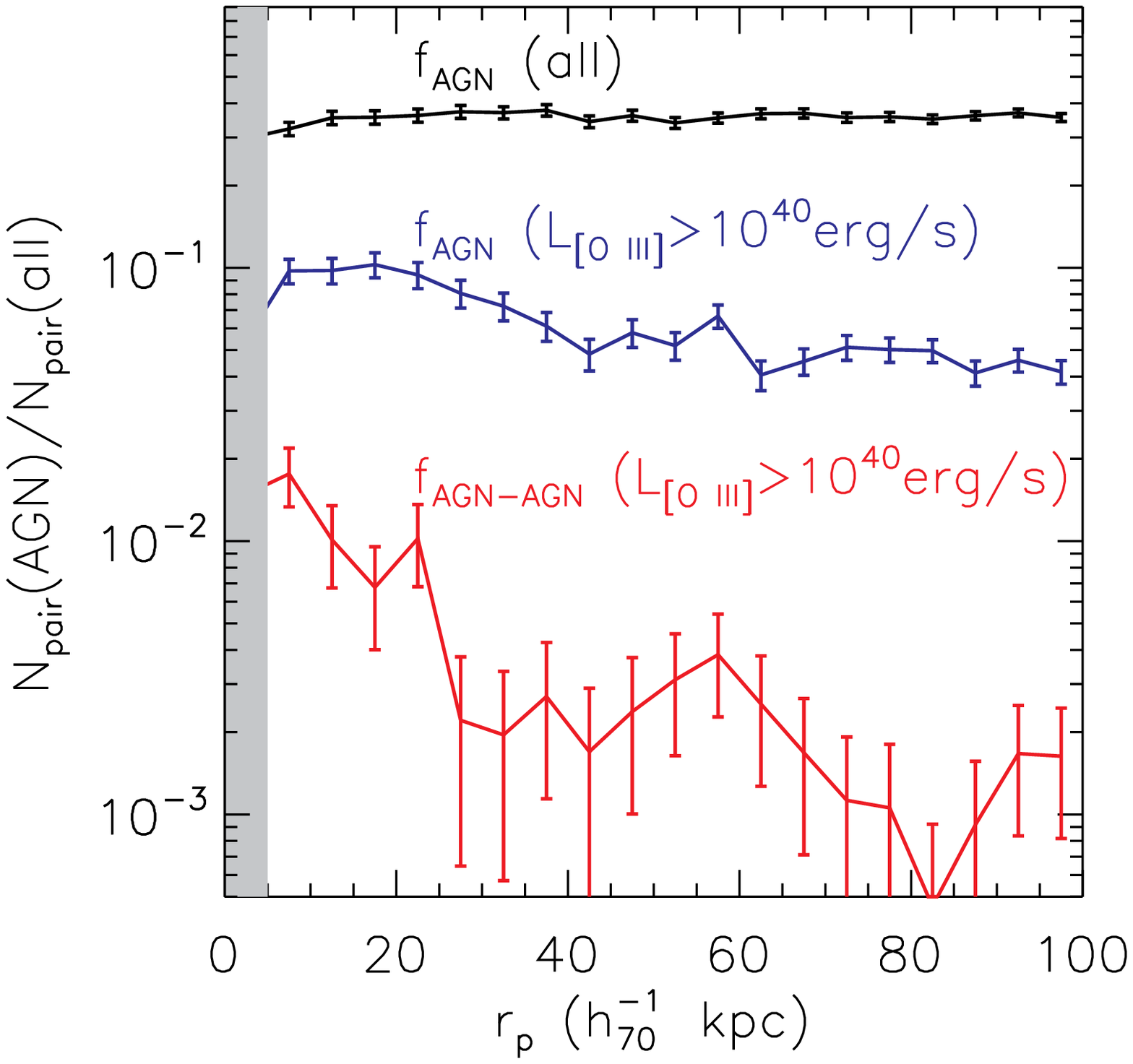}
    \caption{The fraction of galaxy pairs hosting AGNs (either single or double) among all
    galaxy pairs $f_{{\rm AGN}}$ (on a logarithmic scale) as a function of separation.
    $f_{{\rm AGN}}$ shows no significant dependence on $r_p$ from $\sim100$
    \hseventy\ kpc to $\gtrsim 5$ \hseventy\ kpc for all AGNs (shown in black),
    consistent with the clustering results of \citet{li06}.
    On the other hand, for the stronger AGNs
    (e.g., \loiii\ $>10^{40}$ erg s$^{-1}$ corresponding to
    $L_{{\rm Bol}}>10^{43.5}$ erg s$^{-1}$), $f_{{\rm AGN}}$
    rises prominently with decreasing $r_p$ for $r_p \lesssim 30$
    \hseventy\ kpc (shown in blue). The fraction of galaxy pairs
    with both members hosting stronger AGNs $f_{{\rm AGN}-{\rm AGN}}$
    also rises with decreasing $r_p$ for $r_p \lesssim 30$
    \hseventy\ kpc (shown in red).
    Error bars indicate Poisson uncertainties.
    }
    \label{fig:agnfrac}
\end{figure}

We discuss several reasons why the causal link between galaxy
interactions and AGN activity has been controversial. First,
the merger-AGN connection seems to depend sensitively on AGN
luminosity. For example, \citet{serber06} analyzed $\sim10^7$
close neighbors of $\sim2,000$ SDSS quasars at $z<0.4$ with
median $L_{{\rm Bol}}\sim10^{45.5}$ erg s$^{-1}$ and found that
they reside in higher local ($<100$ kpc) over-density regions
than do normal $L^{\ast}$ galaxies. In a much lower AGN
luminosity regime (median $L_{{\rm Bol}}\sim10^{43.0}$ erg
s$^{-1}$), \citet{li06} studied cross correlations of
$\sim90,000$ narrow-line AGNs with galaxies at $z\sim0.1$ from
the SDSS DR4 \citep{SDSSDR4}, and found that on scales $<70$
kpc, AGNs cluster marginally more strongly than control samples
of inactive galaxies matched in redshift, stellar mass,
concentration, velocity dispersion, and mean stellar age,
whereas they cluster more weakly than inactive galaxies on
0.1--1 Mpc scales, and cluster similarly on scales larger than
a few Mpc. The AGN luminosity regime we are probing here is
similar to that of \citet{li06}. As we show in Figure
\ref{fig:agnfrac}, the fraction of galaxy pairs hosting either
single or double AGNs\footnote{To make a fair comparison with
the \citet{li06} clustering results, we adopt the same AGN
diagnostic criteria here, i.e., AGNs according to the
\citet{kauffmann03} empirical curve with all four emission
lines detected with S/N$>3$.} among all galaxy pairs $f_{{\rm
AGN}}$ shows no significant dependence on $r_p$ from $\sim100$
\hseventy\ kpc to $\gtrsim 5$ \hseventy\ kpc for all AGNs. This
is consistent with the clustering results of \citet{li06}. On
the other hand, if we consider only the stronger AGNs (e.g.,
\loiii\ $>10^{40}$ erg s$^{-1}$ corresponding to $L_{{\rm
Bol}}>10^{43.5}$ erg s$^{-1}$), $f_{{\rm AGN}}$ rises
prominently with decreasing $r_p$ for $r_p \lesssim 30$
\hseventy\ kpc, with a factor of $\sim2.5\pm0.5$ ($\sim10\pm2$)
increase from $\sim100$ \hseventy\ kpc to $\gtrsim 5$
\hseventy\ kpc for those hosting either single or double (those
hosting double) AGNs. This comparison suggests that the overall
increase we have detected in AGN luminosity/Eddington ratio in
close AGN pairs is a result of tidally enhanced BH accretion
events on top of a background of weak AGN events driven by
secular processes not associated with interactions.

Second, the $r_p$ correlation we detect for AGN luminosity is
not as prominent and significant as for indicators for recent
star formation. Therefore we speculate that small samples of
AGN pairs with tens of objects or even fewer may simply be
insufficient to reveal such a weak correlation.  In addition,
the difference between the overall distributions of AGN
indicators in the pair and control samples is smaller than that
in the tidal sample (Figure \ref{fig:agn}).  Thus, the effects
of tidal interactions on AGN activity may be underestimated due
to contamination by closely separated galaxy pairs that are not
undergoing tidal interactions.  Furthermore, we have drawn
control samples with identical redshift and stellar mass
distributions.  Because galaxy interactions also enhance star
formation activity, their effects on AGN activity may be
underestimated if control samples are matched in color or in
stellar age, as was done in the clustering analysis of
\citet{li06}. Finally, the enhancement in AGN activity we
observe becomes prominent only when studied as a function of
separation and when $r_p$ is sufficiently small ($\lesssim
10$--30 \hseventy\ kpc). As a result, the effect will not be
detectable in samples of galaxy pairs dominated by
larger-separation systems. For example, \citet{darg10b} find
tentative evidence for a slightly higher AGN fraction in
mergers identified for tidal features using SDSS images at
later stages (32\% at $r_p \sim 5$ kpc) than at earlier stages
(23\% at $r_p \sim 13$ kpc), even though these authors find
little overall evidence for increased AGN activity in galaxy
mergers.

We caution that the correlation between AGN luminosity and
projected separation and that of \OIII\ EW between the
interacting components in an AGN pair support the hypothesis
that galaxy tidal interactions enhance AGN activity, but do not
{\it directly} test whether all AGNs are triggered in or
associated with galaxy interactions.  Galaxy interactions may
not be a necessary condition to trigger AGNs -- secular
processes may suffice at least for low-luminosity AGNs; nor is
it a sufficient condition -- there are internal conditions that
need to be fulfilled such as the presence of ample quantities
of gas. Whether galaxy interactions enhance AGN activity, and
whether it is a sufficient condition to trigger AGNs are two
separate issues.

\subsubsection{Is There A Time Delay between the Tidally-induced AGN and Star Formation
Activity?}

To reconcile the elusive link between AGN activity and galaxy
interactions found by some authors with the definitive
connection between star formation and interactions, it has been
suggested that there is a considerable time delay ($\sim
0.1$--1 Gyr) between the interaction-induced star formation and
AGN activity
\citep[e.g.,][]{sanders88,li08b,ellison08,schawinski10}.  For
our AGN pairs, the values of $r_p$ below which the enhanced AGN
and recent star-formation activity becomes significant are
similar, $\lesssim 10$--30 \hseventy\ kpc. This suggests that
there is no significant time delay between the onset of
interaction-enhanced AGN and star formation activity. However,
as our sample does not probe the peak epoch of
interaction-induced activity, which is likely at later stages
with much smaller nuclear separations than is probed by our
sample, we cannot rule out a considerable time delay between
the peak or overall average epochs of tidally induced star
formation and AGN activity. A statistical sample of AGN pairs
with smaller separations is needed to better understand the
tidal effects close to the peak epoch of the induced activity.
A complementary approach to identifying AGN pairs
\citep{liu10b,shen10b,mcgurk11,fu11} based on the selection of
AGNs with double-peaked narrow emission lines
\citep[e.g.,][]{heckman81,zhou04,comerford08,wang09,liu10,smith09}
combined with follow-up spatially resolved near-IR (NIR)
imaging and optical/NIR spectroscopy is more sensitive to
smaller separation pairs ($\lesssim$ a few kpc).

\subsection{Comparison with Simulations of Galaxy Mergers}\label{subsec:comparesim}

Our results are in broad agreement with the picture in which
galaxy-galaxy interactions induce substantial gas inflows which
in turn promote star formation and AGN activity
\citep[e.g.,][]{noguchi86,hernquist89,mihos96,perez06},
although our sample only probes the low-luminosity regime.
Current simulations cannot resolve the scales relevant for star
formation, AGN feeding, and stellar feedback, and thus employ
simplified recipes and assumptions to model star formation and
the radiated AGN luminosity using galaxy properties on larger
scales. Nevertheless, simulations suggest that, depending on
progenitor properties and orbital parameters, the enhancement
of star formation activity due to tidal effects may vary from
levels as low as about twice the isolated case to values
typical of starburst galaxies (20--60 times the isolated case)
and the peak occurs during the interpenetrating phases
\citep[e.g.,][]{dimatteo07}. The enhancement of recent star
formation activity we observe (0.7--0.9$\pm0.2$ dex from $r_p
\gtrsim 100$ to $\sim 5$ \hseventy\ kpc) is at the upper end of
simulation predictions at the first close passage \citep[1 to
4--5 times the isolated case;][]{dimatteo07}.  The enhancement
in AGN luminosity we observe ($\sim 1.0$ dex in \loiii\ from
$r_p \gtrsim 100$ to $\sim 5$ \hseventy\ kpc; Figure
\ref{fig:agn}) is comparable to the prediction from numerical
simulations \citep[e.g., $\sim 1$ dex higher in SMBH accretion
rate just after the first passage than the isolated
case;][]{springel05}.

Numerical simulations of disk galaxies accreting dwarf
satellites suggest that the primary disk galaxy develops a
strong two-armed spiral pattern due to tidal perturbation from
the infalling satellite, which drives substantial gas into the
center and may fuel starburst activity
\citep{mihos94,hernquist95}. On the other hand, \citet{cox08}
find very little, if any, induced star formation for disk
mergers with mass ratios larger than $5:1$, especially for
galaxies with significant bulge components.  The host galaxies
in our samples of AGN pairs typically contain substantial bulge
components, so our detection of enhanced recent star formation
activity in AGN pairs with stellar mass ratios
$M_{\ast,1}/M_{\ast,2}>3$ is in mild conflict with the
prediction of \citet{cox08}. This discrepancy could be due to
differences in the host-galaxy properties such as mass and
structure and interaction orbital properties. Our results are
based on a special subset of all galaxy pairs, i.e., those in
which both galaxies are AGNs. The simulations are not examing
this subset exclusively, so that it is not a fair comparison.
Alternatively it could point to the importance of physical
processes neglected in simulations \citep[e.g., recycled gas
from stellar winds and supernovae;][]{ciotti07}. Simulations
with more tailored host-galaxy and interaction parameters may
provide useful clues to discriminate between these
possibilities.

Simulations of disk galaxies with comparable masses suggest
that the presence of dense central bulges delays the onset of
gravitational instability, and gaseous inflows are weaker and
occur earlier in bulgeless galaxies \citep{mihos96}.  Our
result, that the star-formation enhancement is larger and
becomes significant earlier (i.e., at larger $r_p$) in
late-type than in early-type hosts of AGN pairs, broadly
supports this hypothesis.  The observed difference is small,
however, probably due to the limited range of interaction
phases spanned in our sample. Again, a statistical sample of
AGN pairs with smaller separations is needed to enlarge the
dynamic range probed in separation.

\section{Summary and Conclusions}\label{sec:sum}

We have studied recent star formation and BH accretion
properties in the host galaxies of 1286 AGN pairs at $\bar{z}
\sim 0.08$ and the subset of 256 pairs with unambiguous
morphological tidal features, selected from the SDSS DR7 as
described in Paper I. We have examined the effects of galaxy
tidal interactions on AGN pairs calibrated against control
samples of isolated AGNs matched in redshift and stellar mass.
We have investigated the correlations between recent star
formation and AGN activity and interaction parameters, the
dependence of these correlations on host-galaxy properties, the
correlation of star formation and AGN activity between the
interacting components in each AGN pair, and the correlation
between recent star formation and AGN activity in AGN pairs as
compared to that observed in ordinary AGNs.  Our main findings
are summarized as follows.

\begin{enumerate}

\item[1.] We have found that the strengths of the 4000
    \angstrom\ break and \hdelta\ absorption in the host
    galaxies depend on the projected separation $r_p$ in
    AGN pairs with tidal features. They also depend weakly
    on LOS velocity offset $\Delta v$. AGN pairs with
    smaller $r_p$ ($\Delta v$) have smaller \dnfk\ and
    larger \hdeltaa , which indicate younger mean stellar
    ages and a higher fraction of post-starburst
    populations in the past 0.1 Gyr.  Assuming the
    SSFR-\dnfk\ calibration of \citet{brinchmann04}, the
    inferred median enhancement of specific star formation
    rates is $\sim$0.7--$0.9\pm0.2$ dex from $r_p \geq 100$
    to $\sim 5$ \hseventy\ kpc. The enhancement of recent
    star formation becomes significant ($> 3\sigma$ above
    the isolated case) at $r_p \lesssim 10$--$30$
    \hseventy\ kpc.

\item[2.] We have found that the \OIIIb\ emission-line
    luminosity \loiii\ and the Eddington ratio (using
    \loiii\ to infer bolometric luminosity and adopting
    stellar velocity dispersion as an indicator for black
    hole mass assuming the calibration of
    \citealt{tremaine02}) depend on $r_p$ in AGN pairs with
    tidal features, although the significance is smaller
    and with the trend being less prominent than that seen
    for recent star formation. A weak dependence on $\Delta
    v$ is also detected. AGN pairs with smaller $r_p$
    ($\Delta v$) on average have larger \loiii\ and higher
    Eddington ratios, and the median enhancement of \loiii\
    (Eddington ratio) is $\sim 0.7\pm0.1$ dex ($\sim
    0.5\pm0.1$ dex) from $r_p \geq 100$ to $\sim 5$
    \hseventy\ kpc. The enhancement of AGN activity becomes
    significant at $r_p \lesssim 10$--$30$ \hseventy\ kpc.
    This moderate overall increase we have detected in AGN
    luminosity/Eddington ratio in close AGN pairs is a
    result of tidally enhanced BH accretion events on top
    of a background of weak AGN events driven by secular
    processes not associated with interactions.

\item[3.] We have compared how the effects of tidal
    interactions on AGN pairs vary when we are restricted
    to pairs with tidal features or not.  We have found
    that the dependence of recent star-formation/AGN
    activity on $r_p$ is present in AGN pairs in both
    cases.  On the other hand, the difference between the
    pair and control AGN samples in their distributions of
    the examined BH accretion and host star formation
    properties is much smaller than that observed between
    the tidal and control samples. The discrepancy may be
    explained by contamination from closely separated pairs
    that are not tidally interacting; In addition, some AGN
    pairs without tidal features may indeed be interacting
    but are less affected by tidal effects, such as mergers
    involving more spheroidal components or those on
    retrograde orbits. Nevertheless, our results suggest
    that the effects of galaxy tidal interactions will be
    underestimated using overall properties of galaxy pairs
    selected solely based on projected separation and
    velocity offsets, in particular if the adopted pair
    separation threshold is substantially larger than
    $10$--$30$ \hseventy\ kpc.

\item[4.] We have explored how host-galaxy mass ratio and
    structure regulate the effects of tidal interactions in
    AGN pairs. We have found that, at $5 \lesssim r_p <
    100$ \hseventy\ kpc, the relative enhancement level of
    star-formation activity in interacting AGN pairs with
    stellar mass ratios $M_{\ast,1}/M_{\ast,2}>3$ is
    comparable to those with comparable stellar masses
    ($M_{\ast,1}/M_{\ast,2}<3$) in both components.  While
    the enhancement of AGN activity is observed in AGN
    pairs of stellar mass ratios as high as 10, the
    relative enhancement on average is more prominent in
    the secondary than in the primary AGN (e.g.,
    $1.2\pm0.3$ dex compared to $0.5\pm0.4$ dex in the
    median value of \loiii , or $0.7\pm0.3$ dex compared to
    $0.6\pm0.4$ dex in the median value of Eddington
    ratio). We have detected no significant difference in
    the tidally enhanced activity when we divide the host
    galaxies according to their $r$-band concentration
    index as a proxy for morphology, but there is a
    marginally larger enhancement for late-type hosts. In
    addition, there is tentative evidence that the
    threshold $r_p$ value below which the recent
    star-formation enhancement becomes significant is
    larger in late- than in early-type hosts. These results
    may reflect the stabilization effect of bulges as
    predicted by the \citet{mihos96} galaxy merger
    simulations.

\item[5.] We have observed a statistically significant
    correlation of recent star formation (indicated by
    \hdeltaa\ and \dnfk\ ) between the interacting
    components in an AGN pair (albeit with a large
    scatter), similar to the color correlation known in
    inactive galaxy pairs. Similarly, we have detected a
    statistically significant correlation of AGN activity
    as indicated by \OIIIb\ rest-frame equivalent width
    between the two components in an AGN pair, although the
    scatter is significant.

\item[6.] We have found a correlation between recent star
    formation and AGN luminosity in AGN pairs similar to
    that observed in control samples of ordinary AGNs.
    While AGN pairs follow the same scaling relation of
    control AGNs, their host galaxies exhibit younger mean
    stellar ages, higher starburst fractions, and host more
    powerful AGNs.

\end{enumerate}

In this paper, we have examined the properties of AGN pairs and
compared them with control samples of single AGNs to
characterize the effects of tidal interactions. The effects
that we have found are based on a special population of galaxy
pairs in which the two central SMBHs are both active at the
same time. More generally, galaxy pairs often host single or no
AGNs. Are galaxy pairs hosting double AGNs intrinsically
different from those hosting single or no AGNs (e.g., in terms
of host galaxy, BH properties, and/or local or large-scale
environments), or is the double-AGN phase universal to all
galaxy tidal interactions? Is the triggering of both AGNs
preferential to certain merger configurations, or are the two
BHs just being turned on and off stochastically? In a
subsequent paper, we will compare AGN pairs with galaxy pairs
hosting single or no AGNs to answer these questions.

The analysis of $z\sim0.1$ AGN pairs as a function of
separation or velocity offset as a proxy for merger stage
presented here also needs to be extended to smaller pair
separations (i.e., more advanced stages) and to higher
redshifts, where the effects of mergers and close interactions
are predicted to be even more dramatic. Recently there has been
an increasing number of AGN pairs with $r_p\lesssim$ a few kpc
identified based on the selection of AGNs with double-peaked
narrow emission lines
\citep[e.g.,][]{heckman81,zhou04,comerford08,wang09,liu10,smith09}
combined with follow-up spatially resolved near-IR (NIR)
imaging and optical/NIR spectroscopy
\citep{liu10b,shen10b,mcgurk11,fu11} to discriminate between
double AGNs and single AGNs with complex gas kinematics. This
approach in principle can identify AGN pairs with separations
as small as $\sim100$ pc, i.e., down to the limit set by the
intrinsic size of narrow line regions around single AGNs. At
$z\gtrsim 1$, i.e., close to the peak AGN epoch
\citep[e.g.,][]{marconi04,hasinger05}, there have been studies
on the local environments of single AGNs
\citep[e.g.,][]{georgakakis08,montero09,coil09,silverman09,silverman11}
in the DEEP2 survey \citep{davis03} or the zCOSMOS survey
\citep{lilly07,scoville07} using X-ray identification of AGNs
based on {\it Chandra} observations in the AEGIS
\citep{davis07} or C-COSMOS \citep{elvis09} field, although a
statistical study of AGN pairs similar to the analysis present
in this paper is still hampered by small galaxy sample size
combined with the difficulty of resolving close pairs and of
identifying tidal disturbance features. Future multi-object NIR
spectrograph such as the MOSFIRE/Keck \citep{mclean08} combined
with high resolution and deep imaging capability such as
offered by the {\it HST} or ground-based adaptive optics
facilities will enable observations of galaxy samples with
sufficient statistical power, resolution, and sensitivity to
perform a similar study of AGN pairs at $z\gtrsim 1$.

\acknowledgments

X.L. thanks J. Goodman, J. Gunn, and J. Krolik for helpful
comments on selection effects in the observed correlations. We
thank an anonymous referee for a very careful and helpful
report.  X.L. and M.A.S. acknowledge the support of NSF grant
AST-0707266. Support for the work of X.L. was provided by NASA
through Einstein Postdoctoral Fellowship grant number
PF0-110076 awarded by the {\it Chandra} X-ray Center, which is
operated by the Smithsonian Astrophysical Observatory for NASA
under contract NAS8-03060.. Y.S. acknowledges support from a
Clay Postdoctoral Fellowship through the Smithsonian
Astrophysical Observatory.

Funding for the SDSS and SDSS-II has been provided by the Alfred
P. Sloan Foundation, the Participating Institutions, the National
Science Foundation, the U.S. Department of Energy, the National
Aeronautics and Space Administration, the Japanese Monbukagakusho,
the Max Planck Society, and the Higher Education Funding Council
for England. The SDSS Web Site is http://www.sdss.org/.

The SDSS is managed by the Astrophysical Research Consortium for
the Participating Institutions. The Participating Institutions are
the American Museum of Natural History, Astrophysical Institute
Potsdam, University of Basel, University of Cambridge, Case
Western Reserve University, University of Chicago, Drexel
University, Fermilab, the Institute for Advanced Study, the Japan
Participation Group, Johns Hopkins University, the Joint Institute
for Nuclear Astrophysics, the Kavli Institute for Particle
Astrophysics and Cosmology, the Korean Scientist Group, the
Chinese Academy of Sciences (LAMOST), Los Alamos National
Laboratory, the Max-Planck-Institute for Astronomy (MPIA), the
Max-Planck-Institute for Astrophysics (MPA), New Mexico State
University, Ohio State University, University of Pittsburgh,
University of Portsmouth, Princeton University, the United States
Naval Observatory, and the University of Washington.

Facilities: Sloan



\bibliography{binaryrefs}

\end{document}